\documentstyle[seceq,preprint,mbf]{ptptex}

\def\slash#1{\ooalign{\hfil/\hfil\crcr$#1$}}

\def\XZ{{\mbf Z}}
\def\XP{{\mbf P}}
\def\XQ{{\mbf Q}}
\def\XM{{\mbf M}}
\def\XA{{\mbf A}}
\def\XD{{\mbf D}}
\def\XS{{\mbf S}}
\def\XK{{\mbf K}}
\def\XU{{\mbf U}}
\def\XX{{\mbf X}}
\def\XG{{\mbf G}}
\def\XV{{\mbf V}}
\def\XL{{\mbf L}}
\def\XH{{\mbf H}}

\def\dt{\!\cdot\!}
\def\nn{\nonumber\\}

\def\calA{{\cal A}}
\def\calD{{\cal D}}
\def\calN{{\cal N}}
\def\hatD{{\hat\calD}}

\def\slashD{{\hat{\slash\calD}}}

\def\R{{\rm R}}
\def\L{{\rm L}}
\def\hatR{{\hat R}}

\def\calF{{\cal F}}

\def\calP{{\cal P}}

\def\half{\hbox{\large ${1\over2}$}}
\def\myfrac#1#2{\hbox{\large ${#1\over#2}$}}
\def\T{{\rm T}}
\def\c{{\rm c}}

\def\6#1{{\underline{#1}}}
\def\m6#1{{\underline{#1}\,}}
\def\4#1{{#1}^{(4)}}
\def\5#1{{#1}^{(5)}}
\def\1#1{{#1}^{(1)}}
\newdimen\Tdim
\def\ispan{{\setbox0=\hbox{i}%
\Tdim\ht0\advance\Tdim\dp0\rule[-\dp0]{0pt}{\Tdim}}}
\def\jspan{{\setbox0=\hbox{j}%
\Tdim\ht0\advance\Tdim\dp0\rule[-\dp0]{0pt}{\Tdim}}}
\def\Tspan#1{{\setbox0=\hbox{#1}%
\Tdim\ht0\advance\Tdim\dp0\advance\Tdim.55ex\rule[-\dp0]{0pt}{\Tdim}\box0}}

\def\Span#1{\setbox0=\hbox{$#1$}\rule[-\dp0]{0pt}{\ht0}}

\preprintnumber[3cm]{%<-- [..]: optional width of preprint # column.
KUNS-1775\\March, 2002\\hep-th/0203276}
\title{%        %You can use \\ for explicit line-break
Superconformal Tensor Calculus on an Orbifold in 5D
}
\author{%       %Use \sc for the family name
Taichiro {\sc Kugo}\footnote{E-mail:
kugo@gauge.scphys.kyoto-u.ac.jp}
and Keisuke {\sc Ohashi}\footnote{E-mail:
keisuke@gauge.scphys.kyoto-u.ac.jp}
}

\inst{%         %Affiliation, neglected when [addenda] or [errata]
Department of Physics, Kyoto University, Kyoto 606-8502, Japan
}

\recdate{%      %Editorial Office will fill in this.
April 18, 2002
}

\abst{%
Superconformal tensor calculus on an orbifold $S^1/Z_2$ is given in 
five-dimensional (5D) spacetime. The four-dimensional superconformal Weyl 
multiplet and various matter multiplets are induced on the boundary 
planes from the 5D supermultiplets in the bulk. We identify those 
induced 4D supermultiplets and clarify a general method for coupling the
bulk fields to the matter fields on the boundaries in a superconformal 
invariant manner. }

\begin{document}

\maketitle

\section{Introduction}

It is a very interesting idea that our world may be a 3-brane 
embedded in a higher-dimensional space-time. To investigate seriously 
various possibilities and problems %in particle physics model building 
in this framework of the `brane world scenario',\cite{ref:HW,ref:hier} 
we need an off-shell 
formulation of supergravity in higher dimensions. Having this in mind, 
some groups have been developing supergravity\cite{Zuk12,ref:KO1,ref:KO2} 
and superconformal\cite{ref:BCDWHP,ref:FO} 
tensor calculus in five-dimensional (5D) space-time. 

In this paper we report on superconformal tensor calculus in 5D 
space-time in which the fourth spatial direction, $x^4\equiv y$, is 
compactified on an orbifold $S^1/Z_2$. 
We clarify the 4D superconformal 
multiplets induced on the boundary planes from the 5D bulk fields. In 
particular, we show that the 4D superconformal Weyl multiplet is 
induced on the boundary planes from the 5D bulk Weyl multiplet. 
Similarly to the rigid supersymmetry case,\cite{ref:MP} 
5D bulk Yang-Mills multiplets induce 4D gauge
multiplets on the boundaries if the vector components are assigned even
$Z_2$ parity. A hypermultiplet in 5D bulk produces a 4D chiral multiplet on
the boundary. A linear multiplet in 5D bulk can also yield a 4D chiral 
multiplet on the boundary for a certain $Z_2$ parity assignment, while, 
for the opposite $Z_2$ parity assignment, it does not give 4D {\em 
linear} multiplet but, rather, a general-type multiplet. 

Once we can identify the 4D superconformal multiplets induced 
on the boundary planes, it becomes immediately clear how to couple 
the 4D matter fields on the boundary to the bulk 
supergravity, Yang-Mills and matter fields in a superconformal invariant 
manner. Since the 4D compensating multiplet is also induced from 
the 5D bulk compensating multiplet, we can write down any 4D invariant 
action on the boundary planes using the known invariant action formulas 
of the 4D superconformal tensor 
calculus.\cite{ref:KTvN,ref:FGvN,ref:KU1,ref:KU2,ref:vN,ref:VP} 

Actually this type of tensor calculus on an orbifold was first studied by 
Zucker.\cite{ref:Zuk3} However, his tensor calculus is not a superconformal
but, rather, a supergravity one, in which dilatation $\XD$ and $\XS$-supersymmetry 
are already gauge-fixed. This fact (together with his choice of 
the linear multiplet for the compensator in the 5D bulk) lead to the
inconvenient situation that a quite unfamiliar 
non-minimal version\cite{ref:SW} 
of 4D Poincar\'e supergravity is induced on the boundary planes.
In our case, 4D superconformal symmetry is fully realized on the boundary 
planes. Then, it is clearly seen that the simplest 4D Poincar\'e
supergravity, `old minimal' version,\cite{ref:oldminimal} is induced on the 
boundary planes if the hypermultiplet is chosen as the compensator in
the 5D bulk.

This paper is organized as follows. We first, 
in \S\S 2 and 3, respectively, 
state the 5D and 4D superconformal transformation rules 
of the Weyl multiplets and some matter multiplets. 
Comparing these transformation rules of the 5D bulk multiplets 
with the 4D transformation rules, as done by 
Mirabelli and Peskin\cite{ref:MP} in the rigid supersymmetry case, 
we identify in \S4 
all the basic 4D supermultiplets induced from the bulk field 
on the boundary. Then, in \S5 we identify the 4D compensator induced 
from the bulk hypermultiplet compensator and explain how the 
brane action can be written down generally. This action 
gives superconformal invariant couplings of the bulk 
supergravity, Yang-Mills and matter fields to an arbitrary set of 4D matter 
fields on the boundary. 
Finally, in \S6 we illustrate the procedure for writing invariant actions 
by considering the simplest case, in which the bulk is a pure supergravity 
system with $U(1)_R$ gauged and the boundary planes contain only tension 
terms. For convenience and to facilitate 
the practical use of the present tensor calculus, 
we add three appendices. The notation and 
conventions are briefly explained in Appendix A. 
Explicit expressions for the curvatures 
both in 5D and 4D cases are given in Appendix B. These make manifest 
the structure functions of the superconformal algebras in both cases. 
Previous results for the embedding and invariant action formulas 
in 4D superconformal tensor calculus are given 
in Appendix C in our present notation. 

\section{5D superconformal multiplets}

Tensor calculus for 5D supergravity was first formulated by 
Zucker,\cite{Zuk12} and later 
by the present authors\cite{ref:KO1} in a more complete 
form.\footnote{These are off-shell formulations of 5D supergravity. 
For on-shell formulations of 5D supergravity, which have been known 
for a long time, see Refs.\citen{ref:5Donshell}}
These formulations, 
however, do not contain conformal $S$ supersymmetry (nor dilatation $D$ 
symmetry in the case of the former), which leads to inconvenience when 
considering a general matter-coupled system, because we must carry out 
 very tedious field redefinitions in order to recover the canonical 
Einstein and Rarita-Schwinger terms.\cite{ref:KO2} These tedious 
field redefinitions 
can simply be bypassed by choosing improved gauge fixing conditions 
of $S$ supersymmetry and dilatation $D$ symmetries in the superconformal 
framework.\cite{ref:KU3} 
In view of this, Bergshoeff et al.\cite{ref:BCDWHP} have derived the Weyl 
multiplets in 5D superconformal tensor calculus, 
and almost simultaneously Fujita and Ohashi  
presented the full superconformal tensor calculus including matter 
multiplets and invariant action formulas in a paper\cite{ref:FO} 
that we refer to as I henceforth. We here restate the 
transformation laws of the Weyl and matter multiplets given in I. 
For the purpose of convenience, which emphasis on practical use, 
we here 
list the explicit expressions for the superconformal covariant 
derivatives and curvatures, which were omitted in I.

\subsection{5D Weyl multiplet}

The 5D Weyl multiplet 
consists of 32 Bose plus 32 Fermi fields, 
\begin{eqnarray}
e_\mu{}^a,\quad \psi_\mu^i,\quad V_\mu^{ij},\quad b_\mu,\quad v^{ab},\quad 
        \chi^i,\quad D,
\end{eqnarray}
whose properties are summarized in table \ref{table:5DWeyl}.
We use $a,b,\cdots$ for the local Lorentz indices, 
$\mu,\nu,\cdots$ for the world vector indices and $i,j=1,2$ for $SU(2)$.
\begin{table}[t]
%\begin{wraptable}{l}{\halftext}
\caption{Weyl multiplet in 5D.}
\label{table:5DWeyl}
\begin{center}
\begin{tabular}{ccccc} \hline \hline
    field      & type   & remarks & SU(2) & Weyl-weight    \\ \hline 
%\multicolumn{5}{c}{Weyl multiplet} \\ \hline
\Tspan{$e_\mu{}^a$} &   boson    & f\"unfbein    & \bf{1}    &  $ -1$   \\  
$\psi^i_\mu$  &  fermion  & SU(2)-Majorana & \bf{2}
%&$-\hspace{-1.5mm} \myfrac12$ \\  
&$-\myfrac12$ \\  
\Tspan{$b_\mu$} & boson &  real & \bf{1} & 0 \\
\Tspan{$V^{ij}_\mu$}    &  boson    
& $V_\mu^{ij}=V_\mu^{ji}=(V_{\mu ij})^*$ & \bf{3}&0\\ 
$v_{ab}$&boson& real, antisymmetric &\bf{1}&1 \\
$ \chi^i$  &  fermion  & SU(2)-Majorana & \bf{2}    
%&\hspace{-1.1mm}\myfrac32 \\  
&\myfrac32 \\  
$D$    &  boson    & real & \bf{1} & 2 \\ \hline
\multicolumn{5}{c}{dependent gauge fields} \\ \hline
$\omega_\mu{}^{ab}$ &   boson    & spin connection & \bf{1}    &   0\\
$\phi_\mu^i$ & fermion & SU(2)-Majorana & \bf{2} & \myfrac12\\
$f_\mu{}^a$ & boson & real & \bf{1} & 1\\
\hline 
\end{tabular}
\end{center}
%\end{wraptable}
\end{table}
The first four fields $e_\mu{}^a,\ \psi_\mu^i,\ V_\mu^{ij}$ and $b_\mu$ are 
the gauge fields for `translation' $\XP_a$, supersymmetry $\XQ^i$, 
$SU(2)$ $\XU_{ij}$ and dilatation $\XD$ transformations, respectively.
The other gauge fields,
%in the 5D superconformal algebra $F^2(4)$,
$\omega_\mu{}^{ab}$ for the local Lorentz $\XM_{ab}$, 
$\phi^i_\mu$ for the conformal supersymmetry $\XS^i$ and 
$f_\mu{}^{a}$ for the special conformal boost $\XK_a$, 
are dependent fields given by functions of the above independent 
gauge fields, as a result of the imposition of 
the following constraints on the $\XP_a$, $\XQ^i$ and $\XM_{ab}$ 
curvatures, respectively:
\begin{eqnarray}
\hatR_{\mu\nu}{}^a(P)=0 
\quad &\longrightarrow& \quad 
\omega_\mu{}^{ab}=\omega^0_\mu{}^{ab}
+i(2\bar\psi_\mu\gamma^{[a}\psi^{b]}+\bar\psi^a\gamma_\mu\psi^b)-2e_\mu{}^{[a}b^{b]},\nn
%&&\hspace{10em}
&& \quad \hbox{with}\ \ 
\omega^0_\mu{}^{ab}\equiv-2e^{\nu[a}\partial_{[\mu}e_{\nu]}{}^{b]}
+e^{\rho[a}e^{b]\sigma}e_\mu{}^c\partial_\rho e_{\sigma c}, \nn
\gamma^\nu\hatR^i_{\mu\nu}(Q)=0
\quad &\longrightarrow& \quad 
\phi_\mu^i=\left(-\myfrac13e^a_\mu\gamma^b+\myfrac1{24}\gamma_\mu\gamma^{ab}\right)
\hatR_{ab}^{\prime\, i}(Q),\nn
%e^\nu_b\hatR_{\mu\nu}{}^{ba}(M)=0
\hatR_\mu{}^a(M)=0
\quad &\longrightarrow& \quad 
f_\mu{}^a=
%\myfrac16\hatR'_\mu{}^a(M)-\myfrac1{48}\eta_{ab}\hatR'(M),
\left(\myfrac16\delta_\mu^\nu\delta^a_b
-\myfrac1{48}e_\mu^ae^\nu_b\right)\hatR'_\nu{}^b(M).
\label{eq:5DRconstraints}
\end{eqnarray}
Here $\hatR_\mu{}^a(M)\equiv\hatR_{\mu\nu}{}^{ba}(M)e^\nu_b$, and 
the primes on the curvatures indicate that 
$\hatR_{ab}^{\prime\,i}(Q)=\hatR_{ab}^i(Q)|_{\phi_\mu=0}$ and 
$\hatR'_\mu{}^a(M)=\hatR_\mu{}^a(M)|_{f_\nu{}^b=0}$.
%$\hatR(M)\equiv\hatR_{ab}{}^{ba}(M)$.
%The explicit definitions of the curvatures $\hatR_{\mu\nu}{}^A=(\hatR_{\mu 
%\nu}{}^i(Q),\,\hatR_{\mu\nu}{}^{ab}(M),\dots) $ is given in the
%Appendix.
A constraint-independent treatment for these dependent gauge fields 
were given in I, but here we prefer to impose the constraints 
(\ref{eq:5DRconstraints}) explicitly, since this is 
simpler in practice. 

The full $\XQ, \XS$ and $\XK$ transformation laws of the Weyl
multiplet are given as follows. With
 $\delta\equiv\bar\varepsilon^i\XQ_i+\bar\eta^i\XS_i+\xi_K^a\XK_a\equiv\delta_Q(\varepsilon)+
\delta_S(\eta)+\delta_K(\xi_K^a)$, we have
\begin{eqnarray}
\delta e_\mu{}^a&=&-2i\bar\varepsilon\gamma^a\psi_\mu,\nn
\delta\psi_\mu^i&=&{\cal D}_\mu\varepsilon^i+\myfrac12 v^{ab}\gamma_{\mu ab}\varepsilon^i-\gamma_\mu\eta^i,
%(\calD_\mu\varepsilon^i\equiv\partial_\mu\varepsilon^i+\half b_\mu\varepsilon^i
%-\myfrac14\omega_\mu^{ab}\gamma_{ab}\varepsilon^i-V_\mu^{\ i}{}_j\varepsilon^j), 
\nn
\delta b_\mu&=&-2i\bar\varepsilon\phi_\mu-2i\bar\eta\psi_\mu-2\xi_{K\mu},\nn
\delta V_\mu^{ij}&=&-6i\bar\varepsilon^{(i}\phi^{j)}_\mu 
+4i\bar\varepsilon^{(i}\gamma\dt v\psi^{j)}_\mu 
-\myfrac{i}4\bar\varepsilon^{(i}\gamma_\mu\chi^{j)}+6i\bar\eta^{(i}\psi^{j)}_\mu,\nn
\delta v_{ab}
&=&-\myfrac{i}8\bar\varepsilon\gamma_{ab}\chi 
-\myfrac32 i\bar\varepsilon\hatR_{ab}(Q), \nn
\delta\chi^i&=&D\varepsilon^i-2\gamma^c\gamma^{ab}\varepsilon^i\hatD_av_{bc}+\gamma\dt \hatR(U)^i{}_j\varepsilon^j
-2\gamma^a\varepsilon^i\epsilon _{abcde}v^{bc}v^{de}+4\gamma\dt v\eta^i,\nn
\delta D&=&-i\bar\varepsilon\slashD\chi-8i\bar\varepsilon\hatR_{ab}(Q)v^{ab}+i\bar\eta\chi, \label{eq:5dWtr}
\end{eqnarray}
where the fermion bilinears like $\bar\eta\psi_\mu$, $\bar\varepsilon\gamma_{ab}\chi$, etc. 
with their $SU(2)$ spinor indices suppressed, always represent 
the northwest-southeast contraction $\bar\eta^i\psi_{\mu i}$, 
$\bar\varepsilon^i\gamma_{ab}\chi_i$, etc. The dot product 
$\gamma\dt T$ for a tensor 
$T_{ab\cdots}$ generally represents the contraction $\gamma^{ab\cdots}T_{ab\cdots}$. 
The transformation rules of dependent fields, of course, follow from 
those of independent fields and are found to be
\begin{eqnarray}
\delta\omega_\mu{}^{ab}&=&2i\bar\varepsilon\gamma^{ab}\phi_\mu 
-2i\bar\varepsilon\gamma^{[a}\hatR_\mu{}^{b]}(Q)-i\bar\varepsilon\gamma_\mu\hatR^{ab}(Q) \nn
&&-2i\bar\varepsilon\gamma^{abcd}\psi_\mu v_{cd}-2i\bar\eta\gamma^{ab}\psi_\mu 
-4\xi_K{}^{[a}e_\mu{}^{b]},\nn
\delta\phi_\mu^i&=&
\calD_\mu\eta^i
-\myfrac13\gamma_{\mu bc}\eta\,v^{bc}+\gamma^b\eta\,v_{\mu b} 
-\xi_K^a\gamma_a\psi_\mu\nn
&&{}
+\gamma_a\varepsilon^if_\mu{}^a
-\myfrac3{32}i\bar\varepsilon\psi_\mu\chi^i
-\myfrac3{32}i\bar\varepsilon\gamma^a\psi_\mu\gamma_a\chi^i
+i\bar\varepsilon^{(i}\gamma^{ab}\psi_\mu^{j)}\bigl(\hatR_{abj}(Q)
-\myfrac1{32}\gamma_{ab}\chi_j\bigr) \nn
&&{}+\myfrac13\left(\hatR_{\mu b}{}^i{}_j(U)\gamma^b
-\myfrac18\gamma_\mu\gamma\dt\hatR{}^i{}_j(U)\right)\varepsilon^j \nn
&&{}-\myfrac1{12}e_\mu^a\left(
3\hatD_a\gamma\dt v\varepsilon^i+\gamma_{abcd}\hatD^bv^{cd}\varepsilon^i
+\gamma_{ab}\hatD_cv^{cb}\varepsilon^i
-2\gamma^{bc}\varepsilon^i\hatD_bv_{ca}-3\varepsilon^i\hatD^bv_{ba}\right.\nn
&&\hspace{4.3em}\left.-\gamma_{abcde}\varepsilon^iv^{bc}v^{de}
+4v_{ab}v_{cd}\gamma^{bcd}\varepsilon^i
+16v_{ab}v^{bc}\gamma_c\varepsilon+5v^2\gamma_a\varepsilon^i\right), \nn
\delta f_\mu{}^a&=&\calD_\mu\xi_K^a-2i\bar\eta\gamma^a\phi_\mu 
+\myfrac56i\bar\eta\gamma^{abc}\psi_\mu v_{bc}
+i\bar\eta\gamma_b\psi_\mu v^{ab}+\myfrac12i\bar\eta\hatR_\mu{}^a(Q)\nn
&&-\myfrac56i\bar\varepsilon\gamma^{abc}\phi_\mu v_{bc}
+i\bar\varepsilon\gamma_b\phi_\mu v^{ab}
-\myfrac1{12}i\bar\varepsilon^i\gamma^{abc}\hatR_{bcij}(U)\psi_\mu^j\nn
&&-\myfrac16i\bar\varepsilon\gamma^{abcd}\psi_\mu\hatD_bv_{cd}
-\myfrac{i}2\bar\varepsilon\psi_\mu\hatD_bv^{ab}\nn
&&-\myfrac56i\bar\varepsilon\gamma^a\psi_\mu v^2-\myfrac83i\bar\varepsilon\gamma^b\psi_\mu v_{bc}v^{ca}
+\myfrac16i\bar\varepsilon\gamma^{abcde}\psi_\mu v_{bc}v_{de}\nn
&&+e_\mu{}^b\left(\myfrac12i\bar\varepsilon\slashD \hatR_b{}^a(Q)
-\myfrac12i\bar\varepsilon\gamma^a\hatD^c\hatR_{bc}(Q)
+\myfrac1{12}\delta_b^ai\bar\varepsilon\hatR(Q)\dt v
-\myfrac13i\bar\varepsilon\gamma_{bc}\hatR^a{}_d(Q)v^{cd}
\right.\nn
&&\qquad \qquad \left.{}
-\myfrac12i\bar\varepsilon\gamma^a{}_c\hatR_{bd}(Q)v^{cd}
+\myfrac1{12}i\bar\varepsilon\gamma\dt v\hatR_b{}^a(Q)
+\myfrac12i\bar\varepsilon\hatR_{bc}(Q)v^{ac}
\right)\,.
\label{eq:5ddeptr}
\end{eqnarray}
Here the (unhatted) derivative $\calD_\mu$ is covariant only with 
respect to the {\em homogeneous transformations} 
$\XM_{ab},\XD$ and $\XU^{ij}$ 
(and the $\XG$ transformation for non-singlet fields under the Yang-Mills 
group $G$), while the hatted derivative $\hatD_\mu$ 
denotes the fully superconformal covariant derivative; 
that is, with $h_\mu^A$ 
denoting the gauge fields of the transformation $\XX_A$, we have 
\begin{equation}
%\calD_\mu&=& \partial_\mu-\sum_{A=\XM_{ab},\XD,\,\XU^{ij}(,\XG)}h_\mu^A\XX_A, \nn
%\hatD_\mu&=& \calD_\mu-\sum_{A=\XQ^i,\XS^i,\XK_a}h_\mu^A\XX_A.
\calD_\mu\equiv\partial_\mu-\!\sum_{\XX_A=\XM_{ab},\XD,\,\XU^{ij}(,\XG)}
\!\!\!h_\mu^A\XX_A, 
\qquad  
\hatD_\mu\equiv\calD_\mu-\!\sum_{\XX_A=\XQ^i,\XS^i,\XK_a}\!\!h_\mu^A\XX_A.
\end{equation}
%For the readers' convenience, we give here 
The explicit forms of the curvatures $\hatR_{\mu\nu}{}^{A}
= e_\mu^{\,b}e_\nu^{\,a}[\hatD_a,\,\hatD_b]^{A}$
are given in Appendix B. 
The covariant derivatives appearing in Eqs.~(\ref{eq:5dWtr}) and 
(\ref{eq:5ddeptr}) are given explicitly by
\begin{eqnarray}
\calD_\mu\varepsilon^i&=&\left(\partial_\mu-\myfrac14\omega_\mu{}^{ab}\gamma_{ab}+\half b_\mu\right)
\varepsilon^i-V_\mu^{\ i}{}_j\varepsilon^j, \nn
\calD_\mu\eta^i&=&\left(\partial_\mu-\myfrac14\omega_\mu{}^{ab}\gamma_{ab}-\myfrac12b_\mu 
\right)\eta^i-V_\mu^{\ i}{}_j\eta^j,\nn
\calD_\mu\xi^a_K&=&\left(\partial_\mu-b_\mu\right)\xi^a_K-\omega_\mu{}^{ab}\xi_{Kb}, \nn
\hatD_\mu v_{ab}&=& \partial_\mu v_{ab}
+2\omega_\mu{}_{[a}{}^c v_{b]c}-b_ƒÊv_{ab}  
+\myfrac{i}8\bar\psi_\mu\gamma_{ab}\chi 
+\myfrac32 i\bar\psi_\mu\hatR_{ab}(Q), \nn
\hatD_\mu\chi^i&=&\calD_\mu\chi^i-D\psi_\mu^i+2\gamma^c\gamma^{ab}\psi_\mu^i\hatD_av_{bc}
-\gamma\dt \hatR(U)^i{}_j\psi_\mu^j
+2\gamma^a\psi_\mu^i\epsilon _{abcde}v^{bc}v^{de}-4\gamma\dt v\phi_\mu^i,\nn
\calD_\mu\chi^i&=&\left(\partial_\mu-\myfrac14\omega_\mu{}^{ab}\gamma_{ab}
-\myfrac32b_\mu\right)\chi^i-V_\mu^{\ i}{}_j\chi^j\,.
\end{eqnarray}

\begin{table}[tb]
%\begin{wraptable}{l}{\halftext}
\caption{Matter multiplets in 5D.}
\label{table:5DM}
\begin{center}
\begin{tabular}{ccccc}\hline \hline
field      & type      &  remarks & {\it SU}$(2)$&  Weyl-weight    \\ \hline 
\multicolumn{5}{c}{Vector multiplet} \\ \hline
$W_\mu$      &  boson    & real gauge field   &  \bf{1}    &   0     \\
$M$& boson & real& \bf{1} & 1 \\ 
$\Omega^i$& fermion  &{\it SU}$(2)$-Majorana  & \bf{2} 
%&\hspace{-1mm}\myfrac32 \\  
&\myfrac32 \\  
$Y_{ij}$    &  boson    & $Y^{ij}=Y^{ji}=(Y_{ij})^*$   & \bf{3} & 2 \\ \hline
\multicolumn{5}{c}{Hypermultiplet} \\ \hline
$\calA_i^\alpha{} $     &  boson & 
$\calA^i_\alpha=\epsilon^{ij}\calA_j^\beta\rho_{\beta\alpha}=-(\calA_i^\alpha)^*$ &\bf{2}
%& \hspace{-0.8mm}\myfrac32 \rule[-1mm]{0pt}{6mm} \\  
& $\myfrac32^{\Span{]}}$ \\  
$\zeta^\alpha$    &  fermion  & $\bar\zeta^\alpha\equiv(\zeta_\alpha)^\dagger\gamma_0 = \zeta^{\alpha\T}C$ 
& \bf{1}  & 2 \\ 
${\cal F}_i^\alpha$  &  boson    & $\calF_i^\alpha\equiv M^Z \XZ\calA_i^\alpha$,\,
${\cal F}^i_\alpha=-({\cal F}_i^\alpha)^*$  &  \bf{2}   
%&\hspace{-0.8mm}\myfrac52 \rule[-3mm]{0pt}{3mm}\\ \hline
& $\myfrac52_{\Span{]}}$ \\ \hline
\multicolumn{5}{c}{Linear multiplet} \\ \hline
\Tspan{$L^{ij}$}& boson & $L^{ij}=L^{ji}=(L_{ij})^*$  &  \bf{3}   & 3 \\ 
$\varphi^i$ &  fermion  & {\it SU}$(2)$-Majorana & \bf{2}
%&\hspace{-1mm}\myfrac72 \\  
&\myfrac72 \\  
$E_a$ & boson & real,\quad constrained & \bf{1} & 4 \\ 
$N$ & boson & real & \bf{1} & 4 \\ \hline
\end{tabular}
\end{center}
%\end{wraptable}
\end{table}

\subsection{Matter multiplets in 5D}

We here give the transformation rules for 
three kinds of matter multiplets: vector multiplets, 
hypermultiplets and linear multiplets.
The components of these multiplets and their properties are listed
in Table \ref{table:5DM}. 
\subsubsection{Vector multiplet}
All the component fields of this multiplet are Lie-algebra valued.
For example, the first component scalar $M$ is the matrix $M^\alpha{}_\beta=M^I(t_I)^\alpha 
{}_\beta$, where the $t_I$ are (anti-hermitian) generators of the gauge 
group $G$: \ $[t_I,\,t_J]=-f_{IJ}{}^Kt_K$. 
The $\XQ$ and $\XS$ transformation laws of the vector multiplet are
given by
\begin{eqnarray}
\delta W_\mu&=&-2i\bar\varepsilon\gamma_\mu\Omega+2i\bar\varepsilon\psi_\mu M,\nn
\delta M&=&2i\bar\varepsilon\Omega,\nn
\delta\Omega^i&=&-\myfrac14\gamma^{\mu\nu} \hat F_{\mu\nu}(W)\varepsilon^i
-\myfrac12\slashD M \varepsilon^i+Y^i{}_j\varepsilon^j-M\eta^i,\nn
\delta Y^{ij}&=&2i\bar\varepsilon^{(i}\slashD\Omega^{j)}-i\bar\varepsilon^{(i}\gamma\dt v\Omega^{j)}
-\myfrac{i}4\bar\varepsilon^{(i}\chi^{j)}M
-2ig\bar\varepsilon^{(i}[M,\Omega^{j)}]-2i\bar\eta^{(i}\Omega^{j)},
\end{eqnarray}
where the full covariant field strength $\hat F_{\mu\nu}(W)$ and 
covariant derivatives are given explicitly 
by
\begin{eqnarray}
\hat F_{\mu\nu}(W)&=& 2\partial_{[\mu}W_{\nu]}-g[W_\mu,\,W_\nu]
+4i\bar\psi_{[\mu}\gamma_{\nu]}\Omega-2i\bar\psi_\mu\psi_\nu M,\nn
\hatD_\mu M&=&(\partial_\mu-b_\mu)M-g[W_\mu,M]-2i\bar\psi_\mu\Omega,\nn
\hatD_\mu\Omega^i&=&\calD_\mu\Omega^i+\myfrac14\gamma\dt \hat F(W)\psi_\mu^i
+\myfrac12\slashD M \psi_\mu^i-Y^i{}_j\psi_\mu^j+M\phi_\mu^i.
\end{eqnarray}
Note that the gauge coupling constant $g$ used in this paper is 
a {\em symbolic} notation; it represents different values for different 
factor groups when $G$ is not a simple group.

%%%%%%%%%%%%%%%%%%%%%%%%%%%%%%%%%%%%%%%%%%%%%%%%%%%%
\subsubsection{Hypermultiplet}

The hypermultiplet in 5D consists of scalars $\calA_i^\alpha$,
spinors $\zeta^\alpha$ and auxiliary fields $\calF_i^\alpha$. They carry the
index $\alpha~(=1,2,\cdots,2r)$ corresponding to the representation of the gauge group $G'$,
which is lowered (or raised) with a $G'$-invariant tensor  
$\rho_{\alpha\beta}$ (and $\rho^{\alpha\beta}$ with 
$\rho^{\gamma\alpha}\rho_{\gamma\beta}=\delta_\beta^\alpha$) as 
$\calA_{i\alpha}=\calA_i^\beta\rho_{\beta\alpha}$. This multiplet 
gives an infinite dimensional representation of the central charge gauge 
group $U(1)_Z$, which we regard as a subgroup of the full gauge group 
$G=G'\times U(1)_Z$. 

The $\XQ$ and $\XS$ transformation rules of $\calA_i^\alpha$ and
$\zeta^\alpha$  are given by
\begin{eqnarray}
\delta\calA_i^\alpha&=&2i\bar\varepsilon_i\zeta^\alpha,\nn
\delta\zeta^\alpha&=&\slashD\calA^\alpha_j\varepsilon^j-\gamma\dt v\,\varepsilon^j\calA^\alpha_j
-M_*\calA^\alpha_j\varepsilon^j+3\calA^\alpha_j\eta^j, \nn
\delta\calF_i^\alpha 
%&=&\delta\left(\delta_Z(\alpha)\calA^\alpha_i\right)
%=\left(\delta_Z(\alpha)\delta+\delta_Z(\delta\alpha)\right)\calA^\alpha_i\nn
&=&2i\bar\varepsilon_i(\alpha\XZ\zeta^\alpha)+\myfrac{2i}\alpha\bar\varepsilon\Omega^0\calF_i^\alpha,
\end{eqnarray}
where 
%the covariant derivative $\hatD_\mu\calA_\alpha^i$ is given explicitly by
%\begin{eqnarray}
%\hatD_\mu\calA_\alpha^i=\calD_\mu\calA_\alpha^i-2i\bar\psi_\mu^i\zeta_\alpha, \quad 
%\calD_\mu\calA_\alpha^i=(\partial_\mu-\myfrac32b_\mu)\calA_\alpha^i-V_\mu{}^i{}_j\calA_\alpha^j
%-W_{\mu*}\calA^{i\beta}
%\end{eqnarray}
%and 
$\theta_*= M_*,\,\Omega_*,\,\cdots$ represent the $G$ transformations with 
the parameters $\theta$ including the central charge transformation, 
$\delta_G(\theta)=\delta_{G'}(\theta)+\delta_Z(\theta^0)$; more explicitly, e.g.,
\begin{equation}
M_*\calA_i^\alpha=\delta_{G'}(M)\calA_i^\alpha+\delta_Z(M^0=\alpha)\calA_i^\alpha 
=\sum_{I=1}^ngM^I(t_I)^\alpha{}_\beta\calA_i^\beta+\alpha\XZ\calA_i^\alpha\,.
\label{eq:star}
\end{equation}
Here, $\XZ$ is the generator of the 
$U(1)_Z$ transformation. The $U(1)_Z$ transformation of $\calA^\alpha 
_i$ defines an auxiliary field $\calF^\alpha_i\equiv\alpha\XZ\calA^\alpha_i$, where 
$\alpha$ is the scalar component of the $U(1)_Z$ vector multiplet
$\XV^0=(M^0\equiv\alpha,\,W^0_\mu,\,\Omega^{0i},
\,Y^{0ij})$.\footnote{We used the
notation $A_\mu$ to denote the gravi-photon field $W_\mu^0$ in previous 
papers. However, we here use $W_\mu^0$ instead, 
since $A_\mu$ is used to denote 
the $U(1)$ gauge field of the 4D superconformal group.} 
The $U(1)_Z$ transformations of the other components, 
$\XZ\zeta_\alpha$ and $\XZ\calF^\alpha_i$, are defined by requiring 
%the following conditions.
\begin{eqnarray}
0&=&\slashD\zeta^\alpha+\myfrac12\gamma\dt v\,\zeta^\alpha 
-\myfrac18\chi^i\calA^\alpha_i+M_*\zeta^\alpha-2\Omega^i_*\calA^\alpha_i,\nn
0&=&-\hatD^a\hatD_a\calA^\alpha_i+M_*M_*\calA^\alpha_i\nn
&&+4i\bar\Omega_{i*}\zeta^\alpha-2Y_{ij*}\calA^{\alpha j}-\myfrac{i}4\bar\zeta^\alpha\chi 
+\myfrac18\left(D-2\,v^2\right)\calA^\alpha_i.\label{eq:con.H}
\end{eqnarray} 
Note that $\hatD_\mu\zeta^\alpha$ and $M_*\zeta^\alpha$ contain 
the central charge transformation terms 
$-W_\mu^0\XZ\zeta^\alpha$ and $\alpha\XZ\zeta^\alpha$, 
respectively, and that both $\hatD^a\hatD_a\calA^\alpha_i$ and 
$M_*M_*\calA^\alpha_i$ contain $\XZ\calF^\alpha_i=\alpha\XZ(\XZ\calA^\alpha_i)$. Hence these 
conditions (\ref{eq:con.H}) indeed 
determine the $\XZ\zeta_\alpha$ and $\XZ\calF^\alpha_i$. 

The explicit forms of the covariant derivatives 
$\hatD_\mu\calA^\alpha_i$ and $\hatD_\mu\zeta^\alpha$ are given by
\begin{eqnarray}
\hatD_\mu\calA^\alpha_i &=&
\left(\partial_\mu-\myfrac32b_\mu\right)\calA^\alpha_i-V_{\mu ij}\calA^{\alpha j}
-gW_\mu{}^\alpha{}_\beta\calA^\beta_i
-W_\mu^0\myfrac1{\alpha}\calF^\alpha_i -2i\bar\psi_{\mu i}\zeta^\alpha\,, \nn
\hatD_\mu\zeta^\alpha&=& \calD_\mu\zeta^\alpha-\slashD\calA^\alpha_j\psi_\mu^j
+\gamma\dt v\,\psi_\mu^j\calA^\alpha_j+M_*\calA^\alpha_j\psi_\mu^j
-3\calA^\alpha_j\phi_\mu^j\,, \nn
\hbox{with}&&\calD_\mu\zeta^\alpha= %&=& 
\left(\partial_\mu-\myfrac14\omega_\mu{}^{ab}\gamma_{ab}
-2b_\mu\right)\zeta^\alpha 
-gW_\mu{}^\alpha{}_\beta\zeta^\beta-W_\mu^0\XZ \zeta^\alpha\,.
\end{eqnarray}
For completeness, we also give an explicit form for 
$\hatD^a\hatD_a\calA^\alpha_i$, which can be obtained by using the 
formula (2$\cdot$31) in Ref.~\citen{ref:KO1},
\begin{equation}
\delta(\varepsilon)\hatD_a\phi=
\varepsilon^A\XX_A(\hatD_a\phi)
= \varepsilon^A \hatD_a(\XX_A\phi) - \varepsilon^Af_{aA}{}^{\bar B}\XX_{\bar B}\phi.
\end{equation}
If we note that $\varepsilon^Af_{aA}{}^{\bar B}$ in the last term is
 equal to the terms containing no gauge fields in 
$e^\mu_a\delta(\varepsilon)h_\mu^{\bar B}$ (i.e., terms proportional to the vielbein 
$e_\mu^c$ in $\delta(\varepsilon)h_\mu^{\bar B}$), we easily find
\begin{eqnarray}
\hatD^a\hatD_a\calA^\alpha_i &=&
\left(\partial^a-\myfrac52b^a\right)(\hatD_a\calA^\alpha_i)
-e^{\mu a}\omega_{\mu ab}(\hatD^b\calA^\alpha_i)
-V^a{}_{ij}\hatD_a\calA^{\alpha j}
-gW^a{}^\alpha{}_\beta\hatD_a\calA^\beta_i \nn
&&{}-W^{0a}\hatD_a(\myfrac1{\alpha}\calF^\alpha_i)
-2i\bar\psi^a_{i}\hatD_a\zeta^\alpha 
-3e^\mu_af_\mu{}^a\calA^\alpha_i
-\myfrac14i\bar\psi_{a(i}\gamma^a\chi_{j)}\calA^{\alpha j} \nn
&&{}-2ig\bar\psi_a\gamma^a\Omega^\alpha{}_\beta\calA^\beta_i 
-2i\bar\psi_a\gamma^a\Omega^0\myfrac1{\alpha}\calF^\alpha_i 
-i\bar\psi^a_i\gamma_{abc}\zeta^\alpha v^{bc}
-2i\bar\phi^a_i\gamma_a\zeta^\alpha\,.\qquad \qquad 
\end{eqnarray}

%%%%%%%%%%%%%%%%%%%%%%%%%%%%%%%%%%%%%%%%%%%%%%%%%%%%%%%%
\subsubsection{Linear multiplet\label{sec:linear}}
The linear multiplet consists of the components listed in 
Table \ref{table:5DM} and may generally carry a charge of
the gauge group $G$. 

The $\XQ$ and $\XS$ transformation laws of the linear multiplet are
given by
\begin{eqnarray}
\delta L^{ij}&=&2i\bar\varepsilon^{(i}\varphi ^{j)},	\nn
\delta\varphi ^i&=&-\slashD L^{ij}\varepsilon_j+\myfrac12\gamma^a\varepsilon^iE_a
+\myfrac12\varepsilon^iN
+2\gamma\dt v\varepsilon_jL^{ij}+M_*L^{ij}\varepsilon_j-6L^{ij}\eta_j, \nn
\delta E^a&=&2i\bar\varepsilon\gamma^{ab}\hatD_b\varphi 
-2i\bar\varepsilon\gamma^{abc}\varphi v_{bc}
+6i\bar\varepsilon\gamma_b\varphi v^{ab}
+2i\bar\varepsilon\gamma^aM_*\varphi -4i\bar\varepsilon^i\gamma^a\Omega^j_*L_{ij}
-8i\bar\eta\gamma_a\varphi, \nn
\delta N&=&-2i\bar\varepsilon\slashD\varphi -3i\bar\varepsilon\gamma\dt v\varphi 
+\myfrac12i\bar\varepsilon^i\chi^jL_{ij}
+4i\bar\varepsilon^{(i}\Omega^{j)}_*L_{ij}-6i\bar\eta\varphi, \label{eq:trf.L}
\end{eqnarray}
with $\theta_*$ defined above in Eq.~(\ref{eq:star}). 
The closure of the algebra demands that $E^a$ satisfy the following 
$Q$- and $S$-invariant constraint:
\begin{equation}
\hatD_aE^a+M_*N+4i\bar\Omega_*\varphi
+2Y^{ij}_*L_{ij}=0.
\label{eq:Con.E}
\end{equation}

\section{4D superconformal tensor calculus}

The $N=1$ 4D formulation of 
superconformal tensor calculus has been known for a long 
time.\cite{ref:KTvN,ref:FGvN,ref:KU1,ref:KU2,ref:vN,ref:VP} 
Here we cite the results, mainly following
Kugo and Uehara\cite{ref:KU1,ref:KU2}
in the present notation. However, strangely enough, the transformation 
rules for the multiplets that carry the gauge group charges have not previously been
given in the literature. Our expressions given here are also valid for 
such cases. 

\subsection{Weyl multiplet}

The 4D superconformal group consists of the usual bosonic conformal 
transformations, $\XP_a$, $\XM_{ab}$, $\XD$ and $\XK_a$, plus a bosonic 
$U(1)$ symmetry $\XA$ and fermionic Majorana $\XQ$ and $\XS$ 
supersymmetries. For simplicity of notation, we use the same symbols 
for the gauge fields, curvatures, etc., in 4D as in 5D, although 
they, of course, denote different quantities. From this point, the 
world vector indices $\mu,\nu,\cdots$ and Lorentz indices $a,b,\cdots$ are considered to run 
%vector indices $\mu,\nu,\cdots$ and $a,b,\cdots$ run 
only over $0,1,2$ and $3$. We attach a 
superscript indicating the dimensions, ``(4)'' or ``(5)'', 
when the distinction is 
relevant. The Weyl multiplet in 4D consists of 12 Bose plus 12 Fermi 
gauge fields and no `matter' fields:
\begin{eqnarray}
e_\mu{}^a,\quad \psi_\mu,\quad A_\mu,\quad b_\mu,
\end{eqnarray}
where $A_\mu$ is the gauge field for the $U(1)$ transformation 
$\XA$.
In this 4D case, the $\XM_{ab},\,\XS$ and $\XK_a$ gauge fields $\omega_\mu 
{}^{ab},\,\phi_\mu$ and $f_\mu{}^a$ are also dependent fields, as
stipulated by the usual constraints,
\begin{eqnarray}
\hatR_{ab}{}^c(P)=0,\qquad \gamma^b\hatR_{ab}(Q)=0, \qquad 
%(\leftrightarrow \gamma_{[a}\hatR_{bc]}(Q)=0,) &\nn &
\hatR_{ab}(M)-\myfrac12\widetilde{\hatR}_{ab}(A)=0,
\label{eq:4DRconstraints}	
\end{eqnarray}
where the tilde denotes the dual tensor 
$\widetilde{F}_{ab}\equiv\epsilon _{abcd}F^{cd}/2$.
The solution for the spin connection $\omega_\mu^{ab}$ to the first 
constraint takes the same form as that in 5D given in 
Eq.~(\ref{eq:5DRconstraints}). 
The solutions for $\phi_\mu$ and $f_\mu{}^a$ to the latter two constraints 
have coefficients that differ slightly from  those in 5D, given in
Eq.~(\ref{eq:5DRconstraints}), and are given by
\begin{eqnarray}
\phi_\mu&=&-\myfrac{i}3\gamma^a\hatR'_{\mu a}(Q)
+\myfrac{i}{12}\gamma_{\mu ab}\hatR^{'ab}(Q), \nn
f_\mu{}^a&=&\myfrac14\hatR'_\mu{}^a(M)
-\myfrac18\widetilde{\hatR}_\mu{}^a(A)-\myfrac1{24}e_\mu{}^a\hatR'(M)\,.
\end{eqnarray}  
The $\XQ,\,\XS,\,\XK_a$ and $\XA$ transformation laws of the gauge
fields are given as follows. 
With $\delta=\delta_Q(\varepsilon)+\delta_S(\eta)+\delta_K(\xi_K^a)+\delta_A(\theta)$,
\begin{eqnarray}
\delta e_\mu{}^a&=&-2i\bar\varepsilon\gamma^a\psi_\mu,\nn
\delta\psi_\mu&=&\calD_\mu\varepsilon+i\gamma_\mu\eta+\myfrac34\theta i\gamma_5\psi_\mu,\nn
\delta b_\mu&=&-2\bar\varepsilon\phi_\mu+2\bar\eta\psi_\mu-2\xi_{K\mu},\nn
\delta A_\mu&=&4i\bar\varepsilon\gamma_5\phi_\mu-4i\bar\eta\gamma_5\psi_\mu+\partial_\mu\theta,\nn
\delta\omega_\mu{}^{ab}&=&2\bar\varepsilon\gamma^{ab}\phi_\mu 
-2i\bar\varepsilon\gamma_\mu\hat R^{ab}(Q)
+2\bar\eta\gamma^{ab}\psi_\mu-4\xi_K^{[a}e_\mu{}^{b]},\nn
\delta\phi_\mu&=&\calD_\mu\eta+i\gamma_a\varepsilon f_\mu{}^a-i\xi_K^a\gamma_a\psi_\mu 
+\myfrac{i}4\gamma^a\varepsilon\,\widetilde{\hatR}_{\mu a}(A)
-\myfrac14\gamma^a\gamma_5\varepsilon\,\hatR_{\mu a}(A)-\myfrac34\theta i\gamma_5\phi_\mu,\nn
\delta f_\mu{}^a&=&\calD_\mu\xi_K^a-2i\bar\eta\gamma^a\phi_\mu 
-i\bar\varepsilon\gamma_\mu\hatD_b\hatR^{ab}(Q),
\label{eq:4DWeyl}
\end{eqnarray}
where the covariant derivatives of 
transformation parameters are defined by 
\begin{eqnarray}
\calD_\mu\varepsilon&=&\left(\partial_\mu-\myfrac14\omega_\mu{}^{ab}\gamma_{ab}+\myfrac12b_\mu-
\myfrac{3}4i\gamma_5A_\mu\right)\varepsilon,\nn
\calD_\mu\eta&=&\left(\partial_\mu-\myfrac14\omega_\mu{}^{ab}\gamma_{ab}-\myfrac12b_\mu+
\myfrac{3}4i\gamma_5A_\mu\right)\eta,\nn
\calD_\mu\xi^a_K&=&\left(\partial_\mu-b_\mu\right)\xi^a_K-\omega_\mu{}^{ab}\xi_{Kb}.
\end{eqnarray}

\subsection{Matter multiplets}
\begin{table}[tb]
%\begin{wraptable}{l}{\halftext}
\caption{Weyl and Matter multiplets in 4D.}
\label{table:4DM}
\begin{center}
\begin{tabular}{ccccc}\hline \hline
field      & type      &  remarks &  Weyl-weight    \\ \hline 
\multicolumn{4}{c}{Wyle multiplet} \\ \hline
$e_\mu{}^a$      &  boson    & real  & $-1$     \\
$\psi_\mu$ & fermion & Majorana &\hspace{-1mm} $-\myfrac12$ \\ 
$A_\mu$ & boson  & real  & $0$ \\  
$b_\mu$ &  boson    & real   & $0$ \\ \hline
\multicolumn{4}{c}{ Complex (real) general multiplet} \\ \hline
$C$ & boson & complex (real)  & $w$ \\
$\zeta$ & fermion & Dirac (Majorana) & $w+\myfrac12$ \\
$H,\,K,\,B_a$ & boson & complex (real) &$ w+1$ \\
$\lambda$ & fermion & Dirac (Majorana) & $w+\myfrac32$ \\
$ D $ &boson & complex (real) & $w+2$\\ \hline
\multicolumn{4}{c}{ gauge multiplet $(w=n=0)$} \\ \hline
$B^g_\mu$ & boson & adjoint rep. & $0$\\
$\lambda^g $ & fermion &Majorana,   adjoint rep. & $\myfrac32$\\
$D^g$ & boson & adjoint rep. & $2$\\ \hline
\multicolumn{4}{c}{ chiral multiplet $(w=n)$} \\ \hline
$\calA$ & boson & complex & $w$\\
$\chi$ & fermion & Majorana & $w+\myfrac12$\\
$\calF$ & boson & complex & $w+1$\\ \hline
\multicolumn{4}{c}{ real linear multiplet $(w=2,\,n=0)$} \\ \hline
$C^L$ & boson & real & $2$\\
$\zeta^L$ &fermion & Majorana & $\myfrac52$\\
$B_a^L$ & boson & real, constrained & $3$\\ \hline  
\end{tabular}
\end{center}
%\end{wraptable}
\end{table}

\subsubsection{Gauge multiplet}

A multiplet that contains the gauge field of a gauge group $G$ is a
gauge multiplet $[B_\mu^g,\lambda^g,D^g]$. The $\XQ,\,\XS$ and
$\XK$ transformation laws are given by
\begin{eqnarray}
\delta B_\mu^g&=&-i\bar\varepsilon\gamma_\mu\lambda^g,\nn
\delta\lambda^g&=&-\myfrac12\gamma\dt \hat F(B^g)\varepsilon+i\gamma_5\varepsilon D^g+\myfrac34\theta i\gamma_5\lambda^g,\nn
\delta D^g&=&\bar\varepsilon\gamma_5\slashD\lambda^g,
\label{eq:4Dgauge}
\end{eqnarray}
where $\hat F_{\mu\nu}(B^g)\,\left(=\hatR_{\mu\nu}(G)\right)$ is a
super-covariantized field strength given by
\begin{eqnarray}
\hat F_{\mu\nu}(B^g)&=&
2\partial_{[\mu}B^g_{\nu]}-[B_\mu^g,B_\nu^g]+2i\bar\psi_{[\mu}\gamma_{\nu]}\lambda^g,\nn
\hatD_\mu\lambda^g&=&\calD_\mu\lambda^g
+\myfrac12\gamma\dt \hat F(B^g)\psi_\mu-i\gamma_5\psi_\mu D^g,\nn
\hbox{with}
&&\calD_\mu\lambda^g=
(\partial_\mu-\myfrac14\omega_\mu{}^{ab}\gamma_{ab}-\myfrac32b_\mu-\myfrac34A_\mu i\gamma_5)\lambda^g.
\end{eqnarray}  
As is well known in the rigid supersymmetry case, this gauge multiplet is 
embedded into a superfield strength multiplet $W_\alpha$, a chiral 
multiplet with an external spinor index $\alpha$, whose first component 
is $\lambda^g_\alpha$. [See Ref.~\citen{ref:KU2} for discussion of 
superconformal multiplets with external Lorentz indices.]

\subsubsection{Complex (or real) general multiplet $\Phi$}

A maximal unconstrained multiplet whose first component is a complex 
scalar $C$, is called 
a `complex general multiplet'. Its full components are
listed in Table \ref{table:4DM}.
The dilatation and $U(1)$ transformations of the first component $C$ 
define the Weyl and chiral weights, $w$ and $n$, of the multiplet,
% The scalar $C$ may transform under the dilatation and $U(1)$
\begin{eqnarray}
\delta_D(\lambda_D)C=w\lambda_DC\,,\qquad \delta_A(\theta)C=\myfrac{i}2n\theta C\,,
\end{eqnarray}
and the multiplet is characterized by these weights $w$ and $n$.
When the chiral weight vanishes ($n=0$), the complex general 
multiplet decomposes into two irreducible real general multiplets,
whose components are all real or Majorana fields. 
The transformation laws of the complex general multiplet
are given as follows (with those of the real general multiplet
obtained by simply setting $n=0$): 
\begin{eqnarray}
\delta C&=&i\bar\varepsilon\gamma_5\zeta+\myfrac12in\theta C,\nn
\delta\zeta&=&\left(i\gamma_5H-K+i\slash B+\slashD C\gamma_5\right)\varepsilon\nn
&&+2i(n+w\gamma_5)C\eta+\left(\myfrac12in-\myfrac34i\gamma_5\right)\theta\zeta,\nn
\delta H&=&\bar\varepsilon\gamma_5\slashD\zeta+\bar\varepsilon i\gamma_5\lambda-\bar\eta\left((w-2)i\gamma_5+in\right)\zeta 
+\left(\myfrac12inH+\myfrac32K\right)\theta,\nn
\delta K&=&i\bar\varepsilon\slashD\zeta-\bar\varepsilon\lambda-\bar\eta\left((w-2)+n\gamma_5\right)\zeta 
+\left(\myfrac12inK-\myfrac32H\right)\theta,\nn
\delta B_a&=&-\bar\varepsilon\hatD_a\zeta-i\bar\varepsilon\gamma_a\lambda\nn
&&-i\bar\eta\left((w+1)+n\gamma_5\right)\gamma_a\zeta+\myfrac12in\theta B_a+2ni\xi_{Ka}C,\nn
\delta\lambda&=&-\myfrac12\gamma\dt \hat F\varepsilon+i\gamma_5\varepsilon D\nn
&&+\left(i\gamma_5H+K-i\slash B-\slashD C\gamma_5\right)(w+n\gamma_5)\eta\nn
&&+\left(\myfrac12in+\myfrac34i\gamma_5\right)\theta\lambda+i(w+n\gamma_5)\xi_K^a\gamma_a\zeta,\nn
\delta D&=&\bar\varepsilon\gamma_5\slashD\lambda+\myfrac12in\theta D\nn
&&-\bar\eta(w\gamma_5+n)\slashD \zeta-2i\bar\eta(w\gamma_5+n)\lambda 
+2w\xi^a_K\hatD_aC+2ni\xi_K^aB_a,\label{eq:4Dgeneral} 
\end{eqnarray}
where $\hat F_{ab}$ is a field-strength-like quantity given by 
\begin{eqnarray}
\hat F_{ab}&=&2\hatD_{[a}B_{b]}
+\myfrac12\epsilon _{abcd}[\hatD^c,\,\hatD^d]C.
\end{eqnarray}
To this point, this general multiplet has tacitly been assumed to 
carry no extra charges. If it carries charges of the gauge 
group $G$, the transformation rules are slightly modified. 
First, the $G$-covariantization term $-\delta_G(B_\mu^g)$ should also be 
included in the full-covariant derivative $\hatD_\mu$. 
Second, the following terms should be added to the above transformation 
laws (\ref{eq:4Dgeneral}):
\begin{eqnarray}
\Delta[\delta B_a]&=&\bar\varepsilon\gamma_5\gamma_a\lambda^g_*C\,,\nn
\Delta[\delta\lambda]&=&-\varepsilon D^g_*C+\gamma^a\varepsilon\myfrac12\bar\lambda^g_*\gamma_a\zeta 
+\gamma^a\gamma_5\varepsilon\myfrac12\bar\lambda^g_*\gamma_a\gamma_5\zeta\,,\nn
\Delta[\delta D]&=&\bar\varepsilon\gamma_5\gamma^a\lambda^g_*B_a
+i\bar\varepsilon\slashD(\lambda^g_*C)+\bar\varepsilon D^g_*\zeta\,,
\end{eqnarray}
where $\theta_*=(\lambda^g_*,\,D^g_*)$ denotes the $\XG$ transformation, 
$\theta_*\Phi=\delta_G(\theta)\Phi$. 

\subsubsection{Chiral multiplet $\Sigma$}

The chiral multiplet $\Sigma=[\calA,\,
\calP_\R\chi,\,\,\calF]\,(\calP_\R\equiv\myfrac12(1+\gamma_5))$ 
can exist when $w=n$, and the anti-chiral multiplet 
$\Sigma^*=[\calA^*,\,\calP_\L\chi,\,\,\calF^*]$, where 
$\calP_\L\equiv\myfrac12(1-\gamma_5)$, when $w=-n$.
Their embedding into the complex general multiplet is given by
% chiral multiplet $w=n$ 
%$\qquad \Sigma=[\calA=\myfrac12(A+iB),\chi,\calF=\myfrac12(F+iG)]$ \\
\begin{eqnarray}
\Phi(\Sigma)&=&[\calA,\ -i\calP_\R\chi,\ 
-\calF,\ i\calF,\ i\hatD_a\calA,\ -2i\calP_\L\lambda^g_*\calA,
\ -iD_*^g\calA],\nn
\Phi(\Sigma^*)&=&[\calA^*\!,\ i\calP_\L\chi,\ 
-\calF^*\!,\ -i\calF^*\!,\ -i\hatD_a\calA^*\!,
\ 2i\calP_\R\lambda^g_*\calA^*\!,\ iD_*^g\calA^*]\,.
\end{eqnarray}
 The transformation laws of these multiplets can be read from
those of $\Phi$ as follows: 
\begin{eqnarray}
\delta\calA&=&\bar\varepsilon\calP_\R\chi+\myfrac{i}2w\theta\calA,\nn
\delta(\calP_\R\chi)&=&\calP_\R\left(-2i\slashD\calA\varepsilon+2\calF\varepsilon 
-4w\calA\eta+\myfrac{i}2(w-\myfrac32)\theta\chi\right),\nn
\delta\calF&=&-i\bar\varepsilon\slashD(\calP_\R\chi)+2\bar\varepsilon\calP_\L\lambda^g_*\calA
+2(w-1)\bar\eta\calP_\R\chi 
+\myfrac{i}2(w-3)\theta\calF\,.
\end{eqnarray}
\subsubsection{Real linear multiplet $L$}
This multiplet, which is denoted by $L=[C^L,\zeta^L,B_a^L]$,
can exist only in the case that the weight is $w=2,\,n=0$. 
The vector component $B_a^L$ is subject to the constraint
\begin{eqnarray}
0=\hatD^aB^L_a-D_*^gC^L+\bar\lambda_*^g\zeta^L\,.
\end{eqnarray}
Interestingly, this constraint is solvable 
in the case that this multiplet that $\XG$-inert
or the matrix $D_*^g$ is invertible.    
This multiplet is also  embedded into the real general multiplet 
in the form
\begin{eqnarray}
\Phi(L)&=&[C^L,\,\zeta^L,\,0,\,0,\,B^L_a,\,i\slashD\zeta^L,\,
\hatD^a\hatD_aC^L-i\bar\lambda^g_*\gamma_5\zeta^L].
\end{eqnarray}
The transformation laws of the components can also be read from those of 
$\Phi$.
%\begin{eqnarray}
%\delta C^L&=&i\bar\varepsilon\gamma_5\zeta^L+\myfrac12in\theta C\nn
%\delta\zeta^L&=&\slashD C^L\gamma_5\varepsilon+i\slash B^L\varepsilon+4i\gamma_5C^L\eta-\myfrac34i\gamma_5\theta\zeta\nn
%\delta B_a^L&=&\bar\varepsilon\gamma_{ab}\hatD^b\zeta^L+\bar\varepsilon\gamma_5\gamma_a\lambda^g_*C^L-3i\bar\eta\gamma_a\zeta^L
%\end{eqnarray}

\section{Identification of $N=1, d=4$ supermultiplets at the boundary}

We must treat 5D and 4D fields simultaneously from this point. 
We use the vector indices $\mu,\nu,\cdots$ and $a,b,\cdots$ as the  
4D indices running over $0,1,2$ and $3$, and write the fifth component 
as $y$ for world vector and as 4 for the Lorentz vector. For instance, 
a 5D vector is written $(V_\mu,V_y)$ or $(V_a,V_5)$.

From the viewpoint of the four-dimensional boundary plane, 
any supermultiplet in the 5D bulk is reducible to an infinite number of 
supermultiplets of 4D superconformal algebra. We here identify 
all the basic 4D supermultiplets that each contains at least one 
bulk field on the boundary without derivative $\partial_y$ (with respect to 
the transverse direction $y$) as a member.

\subsection{4D Weyl multiplet from a 5D one}

The fields $\phi$ are classified as even and odd fields
under the $Z_2$ parity transformation $y\equiv x^4 \rightarrow-y$. 
The $Z_2$ parity eigenvalue $\Pi(\phi)=\pm1$ is defined by 
\begin{equation}
\phi(-y) = \Pi(\phi) \, \phi(y).
\end{equation}
For $SU(2)$-Majorana spinor fermions $\psi^i$ ($i=1,2$), however,  
the $Z_2$ parity transformation mixes the two components $\psi^1$ 
and $\psi^2$ as\cite{ref:BKVP}
\begin{equation}
\psi^i(-y) = \Pi(\psi)\, \gamma_5 M^i{}_j \psi^j(y), 
%\qquad 
%(\bar\psi^i(-y)=\Pi(\psi) M^i{}_j \bar\psi^j(y)\gamma_5), 
\end{equation}
where $M^i{}_j$ is a $2\times2$ matrix satisfying 
$M^* = -\sigma_2M\sigma_2$, where $M=\sigma_3$ 
in our convention. We therefore define 
the combinations of spinors $\psi^i$
\begin{equation}
\begin{array}{ccl}
\psi_+(y) \hspace{-.7em}&\equiv&\hspace{-.7em} 
\psi_\R^1(y) + \psi_\L^2(y), \\[.8ex]
\psi_-(y) \hspace{-.7em}&\equiv&\hspace{-.7em} 
i\bigl( \psi_\L^1(y) + \psi_\R^2(y)\bigr),
\end{array}
\qquad \qquad \left(\psi_{{\R}\atop{\L}}\equiv{1\pm\gamma_5\over2}\psi\right)
\label{eq:Z2spinor}
\end{equation}
which give the $Z_2$ parity eigenstates
\begin{equation}
\psi_\pm(-y)=\pm\Pi(\psi)\psi_\pm(y),  
%\delta_U(\theta^3)\psi_\pm&=&\pm\theta^3i\gamma_5\psi_\pm 
\end{equation}
also satisfying the 4D Majorana property 
$\bar\psi_\pm\equiv(\psi_\pm)^\dagger\gamma^0=\psi_\pm^{\T}C_4$.

The $Z_2$ parity eigenvalues are assigned to fields by demanding the 
invariance of the action and the consistency of both sides of the 
superconformal transformation rules. We list in Table~\ref{table:2} the 
$Z_2$ parity eigenvalues for the Weyl multiplet fields and $Q$- and 
$S$-transformation parameters $\varepsilon$ and $\eta$,\cite{ref:FKO} 
where the `isovector' notation $\vec t=(t^1,t^2,t^3)$ is used, 
which we generally define for
any symmetric $SU(2)$ tensor $t^{ij}$ [satisfying hermiticity 
$t^{ij}=(t_{ij})^*$] according to 
\begin{equation}
t^i{}_j (= t^{ik}\epsilon_{kj}) \equiv i\vec t \cdot \vec\sigma^i{}_j.
\end{equation}
\begin{table}[tb]
%\begin{wraptable}{l}{\halftext}
\caption{$Z_2$ parity eigenvalues\protect\cite{ref:FKO}}
\label{table:2}
\begin{center}
\begin{tabular}{c|c} \hline\hline
\multicolumn{2}{l}{\hspace{10em}{\em Weyl multiplet}} \\ \hline
$\Pi^{\Span{[}}_{\Span{]}}=+1$ &
\ $e_{\mu}^{a},\ \, e_y^4,\ \, \psi_{\mu+},\ \, \psi_{y-},\ \, 
\varepsilon_+,\ \, \eta_-,\ \, b_{\mu},\ \, V_{\mu}^3,\ \, 
V_y^{1,2},\ \, v^{4a},\ \, \chi_+,\ \, D$  \\ \hline
$\Pi^{\Span{[}}_{\Span{]}}=-1$ &
\ $e_{\mu}^{4},\ \, e_y^a,\ \,\psi_{\mu-},\ \,\psi_{y+},\ \, 
\varepsilon_-,\ \,\eta_+,\ \,b_{y},\ \, V_{y}^3,\ \, V_\mu^{1,2},\ \, 
v^{ab},\ \, \chi_-$  \\ \hline
\multicolumn{2}{l}{\hspace{10em}{\em Vector multiplet $\XV$}} \\ \hline
$\Pi_\XV{}^{\Span{A}}_{\Span{A}}$ 
& $M,\ \ W_y,\ \ Y^{1,2},\ \ \Omega_-$  \\ \hline
$-\Pi_\XV{}^{\Span{A}}_{\Span{A}}$ 
& $W_\mu,\ \ Y^{3},\ \ \Omega_+$  \\ \hline
\multicolumn{2}{l}{\hspace{10em}{\em Linear multiplet $\XL$}} \\ \hline
$\Pi_\XL{}^{\Span{A}}_{\Span{A}}$ 
& $L^{1,2},\ \ N,\ \ E^4,\ \ \varphi_+$  \\ \hline
$-\Pi_\XL{}^{\Span{A}}_{\Span{A}}$ 
& $L^{3},\ \ E^a,\ \ \varphi_-$  \\ \hline
\multicolumn{2}{l}{\hspace{10em}{\em Hypermultiplet $\XH$}} \\ \hline
$\Pi_{\hat\alpha}{}^{\Span{[}}_{\Span{]}}$ & 
$\calA^{2\hat\alpha-1}_{\quad i=1},\ \ \calA^{2\hat\alpha}_{\quad i=2},\ \ 
\calF^{2\hat\alpha-1}_{\quad i=2},\ \ \calF^{2\hat\alpha}_{\quad i=1},\ \ 
\zeta^{\hat\alpha}_-$  \\ \hline
$-\Pi_{\hat\alpha}{}^{\Span{[}}_{\Span{]}}$ & 
$\calA^{2\hat\alpha-1}_{\quad i=2},\ \ \calA^{2\hat\alpha}_{\quad i=1},\ \ 
\calF^{2\hat\alpha-1}_{\quad i=1},\ \ \calF^{2\hat\alpha}_{\quad i=2},\ \ 
\zeta^{\hat\alpha}_+$  \\ \hline
\end{tabular}
\end{center}
%\end{wraptable}
\end{table}
The even parity fields are non-vanishing on the 4D boundary planes at 
$y=0$ and $y=\tilde y$ and can form 4D superconformal multiplets there. 
In four dimensions, the parameters of the $\XQ$ and $\XS$ supersymmetry 
transformations are both 4-component Majoranas. In accordance with this,
half of the parameters of 5D the $\XQ$ and $\XS$ supersymmetries vanish
on the boundaries, with only $\varepsilon_+$ and $\eta_-$, respectively,  
%only half of the parameters of 5D $\XQ$ and $\XS$ supersymmetries  
remaining non-vanishing. These non-vanishing parameters 
are indeed the Majorana spinors. 

First of all, the 4D superconformal Weyl multiplet is induced on the 
boundary planes from the 5D bulk Weyl multiplet, and the multiplet members 
can be identified as follows, comparing the superconformal transformation 
laws in 4D and 5D cases; 
that is, the following 5D fields on the right-hand sides can be seen to 
transform in exactly the same way as the 4D Weyl multiplet obeying 
the superconformal transformation rule (\ref{eq:4DWeyl}). 
(Noting that the 5D fields are 
always understood to be those evaluated at the boundary, 
$y=0$ or $\tilde y$, in the relations between the 4D and 5D cases):
%The fields of the 4D Weyl multiplet are identified as follows:
\begin{eqnarray}
\4e_\mu{}^a&=&e_\mu{}^a, \qquad 
\4\psi_\mu=\psi_{\mu+}, \qquad 
\4b_\mu=b_\mu,\nn
\4\omega_\mu{}^{ab}&=&\omega_\mu{}^{ab}, \qquad 
\4A_\mu=\myfrac43\left(V^3_\mu+v_{\mu4}\right),\nn
\4\phi_\mu&=&\phi_{\mu-}-\gamma_5\gamma^av_{a4}\psi_{\mu+}+\Delta\phi_\mu,\nn
\4f_\mu{}^a&=&f_\mu{}^a
%-\myfrac12i\bar\psi_{\mu+}\gamma_5\hatR^a{}_{4}(Q)_- 
-\bar\psi_{\mu+}\Delta\phi^a
+\Delta f_{\mu}{}^a 
%\hat R^{(5)}_{a4}(Q)_-&=&S_a \qquad (\hbox{def.~of } S_a),\nn
%\phi_a^{(5)}(Q)_-&=&\phi_a^{(4)}(Q)+\myfrac13\tilde S_a\nn
%\phi_4^{(5)}(Q)_+&=&
%\myfrac{i}{24}\gamma_5\gamma\dt \hat R^{(4)}(Q)+\myfrac{i}4\gamma^aS_a\nn
%&&\tilde S_a\equiv\left(\delta_a{}^b-\myfrac14\gamma_a\gamma^b\right)
%\left(i\gamma_5S_b-\myfrac{i}2\gamma^c\hat R^{(4)}_{bc}(Q)\right)
\end{eqnarray}
with $\Delta\phi_\mu$ and $\Delta f_{\mu}{}^a$ given by
\begin{eqnarray}
\Delta\phi_\mu&\equiv& \myfrac12i\gamma_5\hatR_{\mu4}(Q)_-, \nn
\Delta f_{\mu}{}^a
&\equiv&-\myfrac16\epsilon_\mu{}^{abc}
\left(\hatD_bv_{c4}+\half\hatR_{bc}{}^3(V)\right) 
+\myfrac14\hatR_{\mu4}{}^{a4}(M).
\label{eq:deltafi}
\end{eqnarray}
Note, however, that 
%\begin{eqnarray}
%\hat R^{(5)i}_{MN}(Q)&=& 2\calD^{(5)}_{[M}\psi^i_{N]}
%+ v^{KL}\gamma_{KL[M}\psi^i_{N]}-2\gamma_{[M}\phi^i_{N]} \nn
%\hat R^{(4)}_{\mu\nu}(Q)&=& 2\calD^{(4)}_{[\mu}\psi^i_{\nu]}
%-2\gamma_{[\mu}\phi^i_{\nu]} 
%\end{eqnarray}
%where $\phi_\nu$ are {\em not} $\phi^{\rm sol}_\nu$ nor 
%$\phi_\nu(Q)$.
%
the 4D $\XQ$ supersymmetry transformation $\4\delta_Q(\epsilon)$ here
is identified with the linear combination of 5D 
transformations at the boundaries
\begin{equation}
\4\delta_Q(\varepsilon=\varepsilon_+)=\delta_Q(\varepsilon_+)+\delta_S(\gamma_5\gamma^av_{a4}\varepsilon_+)
+\delta_K(\bar\varepsilon_+\Delta\phi^a), 
\end{equation}
and the other 4D superconformal transformations, 
the $U(1)$ transformation $\4\delta_A(\theta)$, 
$\XS$ supersymmetry transformation $\4\delta_S(\eta)$, etc., 
are identified as
\begin{eqnarray}
&&\4\delta_A(\theta=\myfrac43\theta^3)=\delta_U(\theta^3), \quad  \qquad 
\4\delta_S(\eta=\eta_-)=\delta_S(\eta_-), \nn
&&\4\delta_D(\rho)=\delta_D(\rho),\qquad 
\4\delta_M(\lambda^{ab})=\delta_M(\lambda^{ab}),\qquad % \nn
\4\delta_K(\xi_K^a)=\delta_K(\xi_K^a). 
\end{eqnarray}
With these identifications of the fields and superconformal 
transformations, we have the following relation between the 
superconformal covariant derivatives in 4D and 5D: 
\begin{eqnarray}
\4\hatD_a&=&\hatD_a-\delta_{U_3}(v_{a4})
-\delta_S(\Delta\phi_a)-\delta_K(\Delta f_a{}^b). 
\label{eq:Drelation}
\end{eqnarray}
If we regard this equation and 
$[\hatD_a,\,\hatD_b]=-\hatR_{ab}{}^A\XX_A$ as holding in any dimensions, 
we can most easily find the expressions (\ref{eq:deltafi}) for $\Delta\phi_\mu$ 
and $\Delta f_{\mu}{}^a$. Indeed, comparing the coefficients of 
$\XX_A=\XQ^i$, $\XM_{ab}$ and $\XU^3$ on both sides of the commutators 
of Eq.~(\ref{eq:Drelation}), we straightforwardly find the relations 
between the curvatures in the 4D and 5D cases:
\begin{eqnarray}
\4\hatR_{ab}(Q)&=&\hatR_{ab}(Q)_+ 
-2\gamma_{[a}\Delta\phi_{b]}, \nn
\4\hatR_{ab}{}^{cd}(M)&=&
\hatR_{ab}{}^{cd}(M)+8\Delta f_{[a}{}^{[c}\delta^{d]}_{b]}, \nn
\myfrac34\4\hatR_{ab}(A)&=&
\hatR_{ab}{}^3(U)+2\hatD_{[a}v_{b]4}.
\end{eqnarray}
Applying to these relations the constraints on the $\XQ$ and $\XM_{ab}$ 
curvatures in both cases, we immediately find the above 
expressions (\ref{eq:deltafi}) for $\Delta\phi_\mu$ 
and $\Delta f_{\mu}{}^a$.

In addition to this 4D Weyl multiplet, the 5D Weyl multiplet also induces 
a 4D `matter' multiplet. Indeed, the extra dimensional component 
$e_y^4$ of the f\"unfbein is also non-vanishing on the boundaries and 
is $\XS$- and $\XK$-inert, so that it can be the first component of a 
superconformal multiplet.\cite{ref:KU2} \ 
It turns out to be a general multiplet 
${\mbf W}_y$ 
with Weyl and chiral weights $(w,n)=(-1,0)$. The identification of the 
multiplet members is given by
\begin{eqnarray}
{\mbf W}_y&\equiv&(C,\zeta,H,K,B_a,\lambda,D) \nn
&=&
\Bigl(e_y^4,\ \,-2\psi_{y-},\ \,-2V_y^2,\ \,2V_y^1,\ \,-2v_{ay},
%\nn
%&&\quad  
\ \,\myfrac{i}4\gamma_5\chi_+e_y{}^4+2\phi_{y+}
+2\gamma_5\gamma^bv_{b4}\psi_{y-},\nn
&&\qquad \bigl(\myfrac14D-(v_{a4})^2\bigr)e_y{}^4
-2f_y{}^4+\myfrac{i}4\bar\chi_+\gamma_5\psi_{y-}
\Bigr).
\label{eq:Wy}
\end{eqnarray}
These fields transform according to the general multiplet 
transformation rule (\ref{eq:4Dgeneral}), provided that the 
covariant derivatives $\4\hatD_\mu$ appearing there are understood to be 
given by 
\begin{eqnarray}
\4\hatD_\mu e_y^4&=&
e_{\mu a}\omega_y{}^{a4}= 
\4\calD_\mu e_y^4+2i\bar\psi_{\mu+}\gamma_5\psi_{y-}-\partial_ye_\mu^4
,\nn
\4\hatD_\mu\psi_{y-}&=&
\4\calD_\mu\psi_{y-}
+[-V_y^1-i\gamma_5V_y^2-i\gamma^bv_{by}+\half(\4\slashD e_y^4)\gamma_5]\psi_{\mu+}
-ie_y^4\gamma_5\4\phi_\mu-\partial_y\psi_{\mu-} \nn
&=&\hatR_{\mu y}{}(Q)_-+i\gamma_\mu\phi_{y+}+i\gamma_5\gamma^av_{a4}\gamma_\mu\psi_{y-},
\label{eq:covD}
\end{eqnarray}
with the `homogeneous covariant derivative'  
$\4\calD_\mu=\partial_\mu-\delta_M(\omega_\mu^{ab})-\delta_D(b_\mu)-\delta_A(A_\mu)$
covariant only with respect to the homogeneous transformations 
$\XM_{ab},\, \XD$ and $\XA$.
The last terms, $-\partial_ye_\mu^4$ in $\4\hatD_\mu e_y^4$ and 
$-\partial_y\psi_{\mu-}$ in $\4\hatD_\mu\psi_{y-}$,  
are unusual and appear as a result of the fact that 
$e_y^4$ and $\psi_{y-}$ carry strange `new' charges, as we now explain.

Generally, if a 5D local transformation parameter $\Lambda(x,y)$ is 
$Z_2$-odd, it vanishes on the boundary. However, its first 
derivative with respect to $y$, $\partial_y\Lambda(x,y)$, is $Z_2$-even and gives a non-vanishing
4D gauge transformation parameter $\partial_y\Lambda(x,0)\equiv\Lambda^{(1)}(x)$ on the 
boundary. Therefore, for the $Z_2$-odd parts of the 5D superconformal 
transformation parameters, there exist the corresponding 4D gauge 
transformations with parameters given as follows:
\begin{equation}
\begin{array}{clcl}
\xi^y &\hbox{of GC transformation $\XP$} & \rightarrow& 
\1\xi(x)\equiv\partial_y\xi^y(x,0), \\
\varepsilon_- &\hbox{of $\XQ$ supersymmetry} & \rightarrow&  
\1\varepsilon(x)\equiv\partial_y\varepsilon_-(x,0), \\
\theta_1,\ \theta_2 &\hbox{of SU(2) transformation $\XU$} & \rightarrow& 
\1\theta_{1,\,2}(x)\equiv\partial_y\theta_{1,\,2}(x,0), \\
\lambda_a{}^4 &\hbox{of local Lorentz $\XM$} & \rightarrow& 
\1\lambda_a(x)\equiv\partial_y\lambda_a{}^4(x,0), \\
\eta_+ &\hbox{of $\XS$ supersymmetry} & \rightarrow& 
\1\eta(x)\equiv\partial_y\eta_+(x,0), \\
\xi_K^4 &\hbox{of special conformal $\XK$} & \rightarrow& 
\1\xi_K(x)\equiv\partial_y\xi_K^4(x,0). 
\end{array}
\label{eq:Dytrfs}
\end{equation}
The general multiplet ${\mbf W}_y$ in Eq.~(\ref{eq:Wy}) transforms 
non-trivially under these transformations. Under the first $\1\xi$ 
transformation, every member of ${\mbf W}_y$ undergoes a common scale 
transformation,
\begin{eqnarray}
\delta{\mbf W}_y&=&\xi^{(1)}{\mbf W}_y,
\end{eqnarray}
and many members of ${\mbf W}_y$ are shifted inhomogeneously 
as Nambu-Goldstone fields under other transformations:
\begin{eqnarray}
\quad \delta{\mbf W}_y&=&
(0,\ -2\varepsilon^{(1)},\ -2\theta^{(1)}_2,\ 
2\theta^{(1)}_1,\ 0,\ 2\eta^{(1)},\ 
-2\xi_K^{(1)}+\myfrac{i}4\bar\varepsilon^{(1)}\gamma_5\chi_+).
\hspace{1em}
\end{eqnarray}
We find the gauge fields for these transformations to be
\begin{equation}
\begin{array}{ccccc}
E_\mu^{(1)}&\equiv&(e_y{}^4)^{-1}\partial_ye_\mu{}^4 & \hbox{for} & \xi^{(1)}, \\
\psi_\mu^{(1)}&\equiv&\partial_y\psi_{\mu-}-E_\mu{}^{(1)}\psi_{y-} & \hbox{for} & \varepsilon^{(1)}, \\
V_\mu^{(1)1,2}&\equiv&\partial_yV_\mu^{1,2}-E_\mu^{(1)}V_y^{1,2} & \hbox{for} & 
\theta^{(1)}_{1,2} ,\\
\vdots&&\vdots& & \vdots \\
\end{array}
\end{equation}
The last terms 
$-\partial_ye_\mu^4$ in $\4\hatD_\mu e_y^4$ and 
$-\partial_y\psi_{\mu-}$ in $\4\hatD_\mu\psi_{y-}$,
 in Eq.~(\ref{eq:covD}) can be 
understood to be identically the terms that appear as the gauge 
covariantization $-\delta_{\xi^{(1)}}(E_\mu^{(1)})-\delta_{\varepsilon^{(1)}}(\psi_\mu^{(1)})$ 
using these gauge fields $E_\mu^{(1)}$ and $\psi_\mu^{(1)}$.
%\begin{eqnarray}
%\delta E_\mu^{(1)}&=&\partial_\mu\xi^{(1)}-\lambda_\mu^{(1)}(e_y{}^4)^{-1}-2\bar\varepsilon\gamma_5\psi_\mu^{(1)}(e_y{}^4)^{-1}
%+2i\bar\varepsilon^{(1)}\gamma_5\psi_\mu(e_y{}^4)^{-1}\nn
%\end{eqnarray}

Since the general multiplet ${\mbf W}_y$ transforms non-trivially under 
these gauge transformations (\ref{eq:Dytrfs}), the utility of the 
multiplet ${\mbf W}_y$ is rather limited. If we wish to use it in 
constructing 4D invariant actions on the brane with the other 
multiplets, we must satisfy the gauge invariance also under the 
transformations (\ref{eq:Dytrfs}), which seems a non-trivial task.

\subsection{From a vector multiplet}

We define the $Z_2$-parity $\Pi_\XV$ of 
the vector multiplet $\XV=(M,\,W_{\mu,y},\,\Omega^i,\,Y^{ij})$ to be 
that of the first scalar component $M$. The $Z_2$-parity quantum numbers
of the other members are given in Table~\ref{table:2}. 

If a vector multiplet $\XV$ is assigned {\it odd} $Z_2$-parity $\Pi 
_\XV=-1$, then the components $W_\mu$, $\Omega_+$ and $Y^3$ are even and 
non-vanishing on the brane, and it gives a 4D gauge multiplet 
$(B^g_\mu,\,\lambda^g,\,D^g)$ defined in Eq.~(\ref{eq:4Dgauge}) 
with $(w,n)=(0,0)$ with the 
following identification: 
\begin{eqnarray}
(B^g_\mu,\ \lambda^g,\ D^g) = (W_\mu,\ 2\Omega_+,\ 2Y^3-\hatD_4M).
%\pr B_\mu&=&W_\mu, \nn
%\lambda&=& 2\Omega_+ \nn
%D &=& 2Y^3-\hatD_4M. 
\end{eqnarray}
This implies that, if $\Pi_\XV=-1$, the bulk Yang-Mills multiplets can 
also couple to the matter multiplets on the brane as the 4D Yang-Mills 
multiplets. 

If a vector multiplet $\XV$ has {\it even} $Z_2$-parity, $\Pi 
_\XV=+1$, then we can identify the following real general-type multiplet
with weight $(w,n)=(1,0)$, whose first component is $M$:
\begin{eqnarray}
&&\hspace{-3em}(C,\,\zeta,\,H,\,K,\,B_a,\,\lambda,\,D) \nn
&=&\Bigl(M,\ -2i\gamma_5\Omega_-,\ 2Y^1,\ 2Y^2,\ \hat F_{a4}(W)+2v_{a4}M,\nn[-.9ex]
&&\quad -2\hatD_4\Omega_+  %-2i\gamma_5M_*\Omega_+ %+2\phi_4(Q)_+M
+2i\gamma^av_{a4}\Omega_--\myfrac{i}4\gamma_5\chi_+M,\ \nn[-.7ex]
&&\quad \hatD_4^2M-2\hatD_4Y^3-\myfrac14DM
+v^a{}_4(2\hat F_{a4}(W)+v_{a4}M)  %\ \nn[-.7ex]
%&&\hspace{8em} {}+2f_{44}^{(5)}(M)
+\myfrac12\bar\chi_+\Omega_- %-4i\bar\Omega\gamma_5\phi_4^{(5)}(Q)
\Bigr). 
\end{eqnarray}
The field $D$ in the term $-\myfrac14DM$ is the auxiliary field 
$D$ in the 5D Weyl multiplet. 

In the latter case of $\Pi_\XV=+1$, we can construct a 4D chiral multiplet 
with weight $(w,n)=(0,0)$:
\begin{eqnarray}
%(\calA,\,\chi,\,\calF) =
%\left(\half(W_y+ie_y^4M),\ 2\psi_{y-}M+2ie_y^4\gamma_5\Omega_-,\ 
%V_y^{1+i2}M-ie_y^4Y^{1+i2}-\bar\psi_{y-}(1+\gamma_5)\Omega_-\right)
\calA &=&
\half(W_y+ie_y^4M)\,, \nn
\chi&=& 
2\psi_{y-}M+2ie_y^4\gamma_5\Omega_-\,, \nn 
\calF &=&
(V_y^1+iV_y^2)M-ie_y^4(Y^1+iY^2)-\bar\psi_{y-}(1+\gamma_5)\Omega_-\,.
\end{eqnarray}
However, this multiplet is also of limited utility because of its 
non-trivial transformation properties
 under the gauge transformation $\1\Lambda 
\equiv\partial_y\Lambda(x,0)$ as well as the above $\1\xi,\ \1\varepsilon,\ \1\theta_1$ and $\1\theta_2$ 
transformations. 

\subsection{From a hypermultiplet}

A hypermultiplet $\XH^\alpha=(\calA^\alpha_i,\,\zeta^\alpha,\, \calF^\alpha_i)$ 
$(\alpha= 1,2,\cdots,2r)$ generally splits into $r$ pairs 
$(\XH^{2\hat\alpha-1}, \XH^{2\hat\alpha})$ $(\hat\alpha=1,2,\cdots,r)$ 
in the standard representation,\cite{ref:dWLVP} in which 
$\rho_{\alpha\beta}= \epsilon \otimes {\bf 1}_r$. 
Then, the following $2\times2$ matrix of 
$(\calA^{2\hat\alpha-1}_i, \calA^{2\hat\alpha}_i)$ 
for each $\hat\alpha$ 
possesses the same real structure as a 
quaternion ${\mbf q}=
q^0+{\mbf i}q^1+{\mbf j}q^2+{\mbf k}q^3$
mapped to a $2\times2$ matrix: 
%\calA^0_{\hat\alpha}+{\mbf i}\calA^1_{\hat\alpha}
%+{\mbf j}\calA^2_{\hat\alpha}+{\mbf k}\calA^3_{\hat\alpha}$}) 
%can be written as
\begin{equation}
\left(
\begin{array}{cc}
\calA^{2\hat\alpha-1}_{\quad i=1} & \calA^{2\hat\alpha-1}_{\quad i=2} \\[.5ex]
\calA^{2\hat\alpha}_{\quad i=1} & \calA^{2\hat\alpha}_{\quad i=2} 
\end{array}\right)
\quad \leftrightarrow \quad 
q^0{\bf1}_2-i\vec q \cdot\vec\sigma= 
\pmatrix{ q^0-iq^3 & -iq^1-q^2 \cr 
-iq^1+q^2 & q^0+iq^3 \cr}. 
%\pmatrix{\calA^{2\hat\alpha-1}_{\quad i=1} & \calA^{2\hat\alpha-1}_{\quad i=2} \cr
%\calA^{2\hat\alpha}_{\quad i=1} & \calA^{2\hat\alpha}_{\quad i=2} \cr}
%=\calA^0_{\hat\alpha}\,{\bf1}_2 - \sum_{k=1}^3 i\calA^k_{\hat\alpha} \,\sigma_k =
%\pmatrix{\calA^0_{\hat\alpha}-i\calA^3_{\hat\alpha} & 
%-i\calA^1_{\hat\alpha}-\calA^2_{\hat\alpha} \cr
%-i\calA^1_{\hat\alpha}+\calA^2_{\hat\alpha} & 
%\calA^0_{\hat\alpha}+i\calA^3_{\hat\alpha} \cr},
\label{eq:quaternion}
\end{equation}
The matrix element fields of this matrix and those of the similar
matrix $(\calF^{2\hat\alpha-1}{}_i, \calF^{2\hat\alpha}{}_i)$ for the auxiliary 
fields give the $Z_2$ parity eigenstates. For spinors, the 
pair $(\zeta^{2\hat\alpha-1}, \zeta^{2\hat\alpha})$ for each $\hat\alpha$ satisfies the 
same realness condition as the 
$SU(2)$-Majorana spinor $\psi^i=(\psi^1,\psi^2)$, so
that we can define the $Z_2$-parity eigenstate 4D-Majorana spinors $\zeta 
^{\hat\alpha}_{\pm}$ in the same way as in Eq.~(\ref{eq:Z2spinor}); $\zeta^{\hat
\alpha}_+\equiv\zeta^{2\hat\alpha-1}_\R + \zeta^{2\hat\alpha}_\L$ and $\zeta^{\hat\alpha}_-\equiv i\bigl
(\zeta^{2\hat\alpha-1}_\L + \zeta^{2\hat\alpha}_\R\bigr)$.

Then, if the 1-1 component $\calA^{2\hat\alpha-1}_{\quad i=1}$ 
has $Z_2$-parity $\Pi_{\hat\alpha}$, 
the $Z_2$-parity quantum numbers of the other hypermultiplet members 
are given as listed in Table~\ref{table:2}.
For either choice of the $Z_2$ parity assignment $\Pi_{\hat\alpha}=\pm1$, 
we obtain the following 4D chiral multiplet with weight 
$(w,n)=(3/2,3/2)$:
%\begin{eqnarray}
%(\calA,\,\chi,\,\calF)&=& 
%\bigl(\calA_{\hat\alpha}^{0+i3},\ -2\zeta^{\hat\alpha}_-,\ 
%\calF^{1+i2}_{\hat\alpha}+gM\calA^{1+i2}_{\hat\alpha}
%-i\hatD_4\calA^{1+i2}_{\hat\alpha}\bigr)\nn
%(\calA,\,\chi,\,\calF)&=& 
%\bigl(\calA_{\hat\alpha}^{1-i2},\ 2\zeta^{\hat\alpha}_+,\ 
%-\calF^{0-i3}_{\hat\alpha}-gM\calA^{0-i3}_{\hat\alpha}
%+i\hatD_4\calA^{0-i3}_{\hat\alpha}\bigr)
%\hspace{2em}
%\end{eqnarray}
\begin{eqnarray}
(\calA,\,\chi_\R,\,\calF)&=& 
\bigl( \calA^{2\hat\alpha}_{i=2},\ \ \ -2i\zeta^{2\hat\alpha}_\R,\ \ \ 
\ (iM_*\calA+\hatD_4\calA)^{2\hat\alpha}_{i=1}\bigr), \nn
(\calA,\,\chi_\R,\,\calF)&=& 
\bigl(\calA^{2\hat\alpha-1}_{i=2},\ -2i\zeta^{2\hat\alpha-1}_\R,\ 
(iM_*\calA+\hatD_4\calA)^{2\hat\alpha-1}_{i=1}\bigr),
\hspace{2em}
\end{eqnarray}
for $\Pi_{\hat\alpha}=\pm1$, respectively. Since 
$M_*\calA^\alpha_i=%(\delta_G(M)+\delta_Z(\alpha))\calA^\alpha_i=
gM^I(t_I)^\alpha{}_\beta\calA^\beta_i+\calF^\beta_i$, the $\calF$-components of these 
chiral multiplets contain the $\calF^\alpha_i$ components of the hypermultiplet.

%\begin{eqnarray}
%e_{(4)}^{-1}{\cal L}_H^4&=&-2i\calA_\pm M_*\calA_\mp ^*-2\calA_\pm\hatD_4\calA_\mp ^*
%-\myfrac{i}2\bar\chi\calP_+\chi\nn
%&&+2\bar\psi\dt\gamma\calP_+\chi\calA_\pm 
%-2\bar\psi_a\gamma^{ab}\calP_-\psi_b\calA^2_\pm+h.c.
%\end{eqnarray}

\subsection{From a linear multiplet}

The $Z_2$ parity quantum numbers for the linear multiplet 
$\XL=(L^{ij},\,\varphi^i,\,E^a,\,N)$, are listed in Table~\ref{table:2}. 

In the case $\Pi_\XL=+1$, we can identify the following 4D chiral 
multiplet with the weight $(w,n)=(3,3)$ on the brane:
\begin{equation}
(\calA,\,\chi,\,\calF)
=\bigl(-L^1+iL^2,\ 2\varphi _+,\ 
\myfrac12(N+iE_4)-\hatD_4L^3-iM_*L^3\bigr). 
\end{equation}
%where $M_*$ was defined in Eq.~(\ref{eq:star}), 
%denotes the group $\XG$ transformation plus 
%central charge transformation $\XZ$ with parameters $gM^I$ and $M^0\equiv\alpha$; 
%$M_*L=\delta_G(M)L+\delta_Z(\alpha)L=gML+\alpha\XZ L$. 
%invariant action 
%\begin{equation}
%e_{(4)}^{-1}{\cal L}_L^4=N-2\hatD_4L^3-2i\bar\psi\dt\gamma\varphi _+
%+2\bar\psi_a\gamma^{ab}(L^1+i\gamma_5L^2)\psi_b
%\end{equation}
In the case  $\Pi_\XL=-1$, the scalar component $L_3$ is non-vanishing on the 
brane. Since it is $\XS$- and $\XK$-inert and carries Weyl and chiral 
weights $(w,n)=(3,0)$, there is a 4D general-type real multiplet
with weight $(w,n)=(3,0)$ starting with $L_3$. We identify the 
components other than the last $D$ component as
\begin{eqnarray}
&&\hspace{-3em}(C,\,\zeta,\,H,\,K,\,B_a,\,\lambda,\,D)\nn
&=&\bigl(L^3,\ -\varphi _-,\ -M_*L^2+\hatD_4L^1,\ M_*L^1+\hatD_4L^2,\ 
-\myfrac12E_a+2v_{a4}L^3,\ \nn[-.5em]
&&\,\qquad {}-i\slashD^{(4)}\varphi _-+i\gamma_5\hatD_4\varphi _+
+M_*\varphi _+-\gamma_5\gamma^av_{a4}\varphi _- \nn
&&\qquad \qquad \qquad {}-2\Omega_{-*}L^1+2i\gamma_5\Omega_{-*}L^2
-\myfrac{i}4\gamma_5\chi_+L^3,\ \ \cdots\ \bigr).
\end{eqnarray}

\section{Compensator and general brane action}

As is well known, we need not only the Weyl multiplet but also a 
special matter field called a `compensator' in order to derive the 
superconformal invariant supergravity actions. The compensator multiplet
is used to fix the extraneous gauge freedoms of the superconformal 
symmetries, like $\XD$, $\XA$ and $\XS$, as well as to saturate the 
required Weyl and chiral weights when applying the invariant action 
formulas. 

The most common formulation in the 4D case is called `old minimal' 
supergravity,\cite{ref:oldminimal} where the compensator used is a 
chiral multiplet $\Sigma$ with weight $w=n=1$. The general superconformal 
invariant action in 4D is given by
%\begin{eqnarray}
%S_{\rm brane} &=&
%\int d^5x \delta(y)\bigl(
%\bigl[\Sigma\bar \Sigma e^{K(S,\bar S)}\bigr]_D 
%+\bigl[f_{IJ}(S)W^{I\alpha}W^J_\alpha\bigr]_F  \nn
%&&\hspace{5em}+\bigl[\Sigma^3 W(S)\bigr]_F \bigr) \ + 
%\int d^5x \delta(y-\tilde y)\bigl( \hbox{the same} \bigr),
%\label{eq:4Daction}
%\end{eqnarray}
\begin{eqnarray}
S_{\rm brane} &=&
\int d^5x \left[\delta(y)+\delta(y-\tilde y)\right]\nn
&&\qquad \times\bigl(
\bigl[\Sigma\bar \Sigma e^{K(S,\bar S)}\bigr]_D 
+\bigl[f_{IJ}(S)W^{I\alpha}W^J_\alpha\bigr]_F 
% \nn &&\hspace{5em}
+\bigl[\Sigma^3 W(S)\bigr]_F \bigr) ,
\label{eq:4Daction}
\end{eqnarray}
where $K(S, \bar S)$ and $W(S)$ are K\"ahler potential and 
superpotential functions, respectively, and 
$[ \cdots]_D$ and $[ \cdots]_F$ 
represent the D-term and F-term 
invariant action formulas in the 4D superconformal 
tensor calculus,\cite{ref:KU1} explained in Appendix C. 
The quantities $S_i$ are the 4D chiral matter 
multiplets with weight $w=n=0$, and $W^I_\alpha$ denotes the 
superfield strength 
of the 4D Yang-Mills multiplets $V^I$. Both of these chiral and gauge 
multiplets, $S$ and $V$, may 
be genuine 4D multiplets existing solely on the brane or induced 
multiplets on the boundary from the 5D bulk multiplets. 
%We should keep it in mind that this formula (\ref{eq:4Daction}) 
%is superconformal invariant but is not automatically $G$-gauge 
%invariant. 

The 4D Weyl multiplet used in expressing the action formulas in 
Eq.~(\ref{eq:4Daction}) should, of course, be the induced Weyl multiplet
found in the previous section. Since gravity is unique, we cannot add 
a {\em genuine} 4D Weyl multiplet on the brane in addition to the induced 
one. In the same sense, we cannot add 
a {\em genuine} 4D compensator on the brane in addition to that induced 
from the bulk compensator fields.

Let us now identify the 4D compensator induced from the 5D bulk 
compensator. The most useful and common choice of the 5D compensator is 
the hypermultiplet, which we discussed in Ref.~\citen{ref:KO2}. 
Consider the 
simplest case of a single-quaternion compensator ($p=1$) in which the 
hypermultiplet compensator is given by $\XH^a=(\calA^a_i,\,\zeta 
^a,\,\calF^a_i)$ ($a=1,2$). Then, as seen in the previous 
section, this hypermultiplet gives the 4D chiral multiplet 
$\Sigma_{\c}=(\calA_{\c},\,\chi_{\c},\,\calF_{\c})$ with weight $(w,n)=(3/2,3/2)$ 
on the brane, assuming the $Z_2$ parity assignment $\Pi_a=+1$:
\begin{equation}
\Sigma_{\c}: \qquad 
\left\{
\begin{array}{lcl}
\calA_{\c}&=&\calA^{a=2}_{i=2} \\
\chi_{\c\R} &=& -2i\zeta^{a=2}_\R \\
\calF_{\c} &=& 
i\calF^{a=2}_{i=1}+ig(M\calA)^{a=2}_{i=1}
+\hatD_4\calA^{a=2}_{i=1}
\end{array}\right.\,.
\label{eq:compen}
\end{equation}
Since this multiplet $\Sigma_{\c}$ carries Weyl and chiral weights 
$w=n=3/2$, we should identify $\Sigma_{\c}^{2/3}$ with the 4D chiral 
compensator $\Sigma$, with $w=n=1$ induced on the brane. 
Note that, if the 5D superconformal gauges are fixed, for example,  
by the conditions
\begin{eqnarray}
\XD,\ \XU^{ij}&:& \quad \calA^a_i
=\left(\begin{array}{cc}
1 & 0 \\
0 & 1 
\end{array}\right)\,, \nn
\XS^i &:& \quad \zeta^a=0
\label{eq:gaugefix}
\end{eqnarray}  
in the bulk, then, these yield the conditions
%$\Sigma\equiv\Sigma_{\c}^{2/3}=(\calA,\,\chi,\,\calF)$ on the boundary,
\begin{eqnarray}
\XD,\ \XA \ : \quad \calA = 1, \qquad 
\XS \ : \quad \chi=0,
\end{eqnarray}  
on $\Sigma\equiv\Sigma_{\c}^{2/3}\!\!=(\calA,\chi,\calF)$ on the
boundary. These are the same 4D superconformal gauge fixing conditions 
as imposed on the usual chiral compensator in the case of a pure
supergravity system.\cite{ref:KU1}

This brane action (\ref{eq:4Daction}) gives superconformal invariant 
coupling between the 4D matters on the brane and the 5D bulk fields 
through the induced Weyl, Yang-Mills and compensating multiplets.

%\section{The Randall-Sundrum supergravity system}
\section{Altendorfer-Bagger-Nemeschansky approach}

%We first explain how the action of Altendorfer-Bagger-Nemeschansky 
%emerges in our superconformal framework and see why the relation
%between the bulk cosmological constant and the brane tensions of 
%boundary planes is not imposed by the supersymmetry requirement.

We next illustrate how the action is given for the bulk-plus-brane system 
by considering the simplest case, in which the bulk is a pure
supergravity system with $U(1)_R$ gauged, and the brane contains only
the tension term.
This is the system that was first constructed by Altendorfer, Bagger 
and Nemeschansky\cite{ref:ABN} in an attempt to supersymmetrize the 
Randall and Sundrum\cite{ref:RS} scenario. 

In our off-shell formulation, 
the system contains a 5D Weyl multiplet, the hypermultiplet compensator 
$\XH^a=(\calA^a_i,\,\zeta^a,\,\calF^a_i)$ ($a=1,2$) 
and a vector multiplet $\XV_0=(M^0=\alpha,\,W^0_\mu,\,\Omega 
^{0i},\,Y^{0ij})$ coupling to the central charge $\XZ$ of the 
hypermultiplet. 
Note here that this central charge vector multiplet $\XV_0$ 
is simultaneously the $U(1)_R$ gauge multiplet coupling to the 
$U(1)_R$-charge which is in general represented by a $2\times2$ matrix 
$(t_R)^a{}_b$ acting on the group index $a=1,2$ of $\XH^a$:
\begin{equation}
t_{\rm R}= i\vec Q\cdot \vec\sigma=i(Q^1\sigma_1+Q^2\sigma_2+Q^3\sigma_3), \qquad 
|\vec Q\,|=1. 
\label{eq:tR}
\end{equation}
Here, $\XV_0$ can be made to play such a double role if we add a `mass term' 
$m\eta^{ab}{\cal L}_{VL}(\XV_0\cdot\XL(\XH_a,\XH_b))$ to the Lagrangian 
with a symmetric tensor $\eta^{ab}$ related to the $U(1)_R$ generator 
$t_R$ by $g_R(t_R)^{ab}=m
\eta^{ab}/2$. 
(Here, ${\cal L}_{VL}(\XV\cdot\XL)$ is the invariant $V$-$L$ action
formula [Eq.~(4$\cdot$5) in I], and $\XL(\XH_a,\XH'_b)$ is the 
embedding formula [Eq.~(4$\cdot$3) in I] of two 
hypermultiplets $\XH$ and $\XH'$ into a linear multiplet.) The bulk 
Lagrangian is given in the form
\begin{eqnarray}
%&&\hspace{-5em}
%-2{\cal L}_{VL}\bigl(\XV_0\cdot\XL(\XH^a,\XZ\XH_a)\bigr)
%-m{\cal L}_{VL}\bigl(\XV_0\cdot\XL(\XH^a,\eta_a{}^b\XH_b)\bigr)
%+{\cal L}_{VL}\bigl(\XV_0\cdot\XL(-c\XV_0^2)\bigr)
%\nn
%&=& 
-2{\cal L}_{VL}\bigl(\XV_0\cdot
\XL(\XH^a,\XZ\XH_b+\half m\eta_b{}^c\XH_c)\bigr)
+{\cal L}_{VL}\bigl(\XV_0\cdot\XL(-c\XV_0^2)\bigr),
\label{eq:BulkAction}
\end{eqnarray}
where $\XZ\XH^a$ is the central-charge transformed hypermultiplet 
whose first component is $\XZ\calA^a{}_i=\calF^a{}_i/\alpha$, and 
$\XL(\half f_{IJ}\XV^I\XV^J)$ denotes generally the embedding formula 
[Eq.~(4$\cdot$1) in I] of vector multiplets $\XV^I$ into a linear 
multiplet based on the homogeneous quadratic function $f(M)=\half 
f_{IJ}M^IM^J$. The quantity 
${\cal L}_{VL}\bigl(\XV_0\cdot\XL(-c\XV_0^2)\bigr)$ 
here thus corresponds to the choice of $f(\alpha)=-c\alpha^2$ and to the
action with `norm function'\cite{ref:GST} $\calN(\alpha)= c\alpha^3$, 
where $c$ is a 
constant coefficient. The explicit component form of the action 
(\ref{eq:BulkAction}) can be read form the general expression given in 
Ref.~\citen{ref:KO2}. The extraneous gauges are generally fixed by the 
gauge conditions
\begin{equation}
\XD:\ \ {\calN}=1,\qquad \XS:\ \ \Omega^{Ii}\calN_I=0,\qquad 
\XK:\ \ \hatD_a{\calN}=0\,.
\label{eq:gauge-fixing}
\end{equation}
The first $\XD$ gauge-fixing condition $\calN=c\alpha^3=1$ may 
equivalently be replaced by $\calA^a_i\calA^i_a=-2$, thanks to the 
auxiliary field equation $\delta S/\delta D 
\propto\calA^a_i\calA^i_a+2\calN =0$. Similarly, if we use the equation 
of motion $\delta 
S/\delta\chi^i\propto\calA^a_i\zeta_a+\calN_I\Omega^I_i=0$, the 
$\XS$-gauge condition $\calN_I\Omega^I_i=0$ is equivalent to 
$\calA^a_i\zeta_a=0$. Then, imposing also $\calA^a_i\propto\delta^a_i$ 
as the $\XU^{ij}$ gauge-fixing conditions, these conditions reproduce 
the previous gauge-fixing conditions (\ref{eq:gaugefix}). The kinetic 
term $-(1/4)F_{\mu\nu}^2(W^0)$ of the gravi-photon field $W^0_\mu$ has 
the coefficient $-(1/2)\calN(\partial^2\ln\calN/\partial\alpha^2)$ in the 
action,\cite{ref:KO2} so that it is properly normalized if the constant 
$c$ is chosen to satisfy $c\alpha=2/3$. Together with the $\XD$ gauge 
condition $\calN=c\alpha^3=1$, this determines $\alpha=\sqrt{3/2}$ and 
$c=(\sqrt{2/3})^3$. The cosmological constant in the bulk is found to be
$-(8/3)g^2_R\alpha^2=-4g_R^2$.\cite{ref:KO2}

Note here that the consistency of the $U(1)_R$ symmetry with the $Z_2$ 
parity requires $Q^3=0$.\cite{ref:BKVP} 
This can be seen if we consider the two terms in
the covariant derivative,
\begin{equation}
\calD_\mu\calA^a_i = \partial_\mu\calA^a_i-g_RW^0_\mu(t_R)^a{}_b\calA^b_i + \cdots 
\label{eq:covA}
\end{equation}
Recall that the scalar component $M^0=\alpha$ of the central charge vector 
multiplet is $Z_2$-even, so that the vector component $W^0_\mu$ is 
$Z_2$-odd, as seen in Table~\ref{table:2}. Since $W^0_\mu$ is $Z_2$-odd and 
$\calA^{a=1}_i$ and $\calA^{a=2}_i$ carry opposite $Z_2$-parity, the 
first and the second terms in Eq.~(\ref{eq:covA}) can have the same 
$Z_2$ even-oddness if and only if $t_R$ possesses only off-diagonal 
components. In other words,
 $t_{\rm R}= i(Q^1\sigma_1+Q^2\sigma_2)$ with no $\sigma_3$ 
component. (If the coupling constant $g_R$ were lifted to the $Z_2$-odd 
field, as in the Gherghetta-Pomarol\cite{ref:GP} and 
Falkowski-Lalak-Pokorski (GPFLP) 
approach,\cite{ref:GP,ref:FLP,ref:BKVP} 
then the $g_RW^0_\mu$ part would become $Z_2$-even and 
%so the conclusion becomes opposite; namely, 
$t_R$ would be diagonal so that $t_R=i\sigma_3$.)\ \ 
After the $SU(2)$ gauge is fixed in the bulk by the condition 
$\calA^a_i\propto\delta^a_i$, the $U(1)_R$ gauge transformation is modified into 
the combined (diagonal) $U(1)$ transformation of the original $U(1)_R$ 
and $SU(2)$; e.g., $\delta_{U(1)_R}(\theta)\calF^a{}_i = i\theta[\,Q^1\sigma_1+Q^2\sigma 
_2,\ \calF\,]^a{}_i$. However, the $SU(2)$ symmetry is {\em explicitly} broken 
by the $Z_2$ parity assignment down to $U(1)$ with the generator $\sigma_3$. 
This breaking is manifest only at the boundaries, since $SU(2)$ is
a local symmetry, while the $Z_2$ parity transformation relates the fields at
$y$ only with those at $-y$. Since the $U(1)$ transformation of the 
generator $Q^1\sigma_1+Q^2\sigma_2$ in $SU(2)$ is broken, the $U(1)_R$ 
symmetry is {\em explicitly} broken in this 
Altendorfer-Bagger-Nemeschansky approach. 

The brane tension terms are given by
\begin{equation}
S_{\rm brane}=
\int d^5x\bigl(\Lambda_1\delta(y)+\Lambda_2\delta(y-\tilde y)\bigr)\,\bigl[\Sigma^3=\Sigma_{\c}^2\bigr]_F\,.
%\int d^5x\bigl(\delta(y)\,\bigl[\Lambda_1\Sigma_{\c}^2\bigr]_F + 
%\delta(y-\tilde y)\,\bigl[\Lambda_2\Sigma_{\c}^2\bigr]_F \bigr),
\label{eq:braneaction}
\end{equation} 
Here $\Lambda_1$ and $\Lambda_2$ are assumed to be real for simplicity and 
%(generally complex) constants.
%This is invariant under the usual superconformal transformations.  
%However, this action formula does not automatically gives a
%$G$ gauge invariant. Indeed 
the F-term action formula (\ref{eq:Fformula}) reads, for 
our chiral compensator $\Sigma_{\c}=(\calA_{\c}=1,\  \chi_{\c}=0,\  \calF_{\c})$, as
\begin{equation}
\bigl[\Sigma_{\c}^2\bigr]_F = e_4 \bigl(
2(\calF_{\c}+\calF^*_{\c}) 
- 2\bar\psi_{\mu}\gamma^{{\mu}{\nu}}\psi_{\nu}\bigr),
\end{equation}
where $e_4$ is the four dimensional determinant of the vierbein,
$e_4=e/e_y^4=e\cdot e^y_4$.  Note that now
\begin{eqnarray}
\hatD_4\calA^a{}_i 
&=& (\partial_4-{3\over2}b_4)\calA^a{}_i 
-{V_4}_{ij}\calA^{aj} -{W^0_4\over\alpha}\calF^a{}_i -2i\bar\psi_i\zeta^a \nn
\rightarrow&&\hatD_4\calA^{a=2}{}_{i=1}=  
 -{W^0_4\over\alpha}\widetilde\calF^2{}_1
+(V_4)^2{}_1-g_\R W^0_4(t_{\rm R})^2{}_1,
\end{eqnarray}
where we have used $\calA^a_i=\delta^a_i$ and $\zeta^a=0$, 
by the superconformal gauge fixing
(\ref{eq:gaugefix}), and $\widetilde\calF^a{}_i$ is defined by
\begin{equation}
\widetilde\calF^a{}_i\equiv 
\calF^a{}_i-\half m\alpha\eta^a{}_b\calA^b{}_i
=\calF^a{}_i-g_R\alpha(t_R)^a{}_b\calA^b{}_i.
\end{equation}
This field $\widetilde\calF^a{}_i$ vanishes in the 
absence of the brane term $\Lambda_1=\Lambda_2=0$. 
%and using $b_4=0$ on the boundary ($\Pi(b_4)=-1$), 
Then we find the real part of the $F$-component 
$\calF_{\c}$ of $\Sigma_{\c}$ in (\ref{eq:compen}) as
\begin{eqnarray}
{\rm Re}\calF_{\c} 
&=&{\rm Re}\left\{ 
i(1+i{W^0_4\over\alpha})\widetilde\calF^2{}_1 +ig_R\alpha(t_R)^2{}_1
+(V_4-g_\R W^0_4 t_{\rm R})^2{}_1
\right\} \nn
&=&
\widetilde\calF^1-{W^0_4\over\alpha}\widetilde\calF^2 -g_R\alpha Q^1
-\widetilde{V}_4^2. 
\end{eqnarray}
Here $Q^1$ is the first component of the direction vector 
$\vec Q\equiv(Q^1,Q^2,Q^3)$ of the $U(1)_{\rm R}$ generator (\ref{eq:tR}) 
in $SU(2)$, 
$\calF^1$ and $\calF^2$ are the 1 and 2 components of the `quaternion' 
$\calF^a{}_i$  $(\calF^a{}_i= 
\calF^0{\bf1}_2-i\calF^1\sigma_1-i\calF^2\sigma_2-i\calF^3\sigma_3)$,
and the quantities $\widetilde{V}_4^k$ are defined by
\begin{equation}
(\widetilde V_4)^i{}_j \equiv(V_4 - g_\R W^0_4 t_{\rm R})^i{}_j 
\equiv\sum_{k=1}^3i\widetilde{V}_4^k(\sigma_k)^i{}_j\,.
\end{equation}

Since the auxiliary fields $\widetilde\calF^1$, $\widetilde\calF^2$ and 
$\widetilde{V}_4^2$ appear 
in the bulk action in the form
\begin{equation}
e\left[ -2\bigl(1+{(W^0_4)^2-W^{0\mu}W^0_\mu\over\alpha^2}\bigr)
\bigl((\widetilde\calF^1)^2+(\widetilde\calF^2)^2\bigr)
+2(\widetilde{V}_4^2)^2 \right]
\label{eq:FVinbulk}
\end{equation}
with opposite signs and $W^0_\mu$ vanishes on the brane ($\Pi(W^0_\mu)=-1$), 
the solution to these auxiliary field equations of motion are given by
\begin{eqnarray}
\widetilde\calF^1&=&\bigl(1+{(W^0_4)^2\over\alpha^2}\bigr)^{-1}
e^y_4(\Lambda_1 \delta(y)+\Lambda_2\delta(y-\tilde y)), \nn
\widetilde\calF^2&=&\bigl(1+{(W^0_4)^2\over\alpha^2}\bigr)^{-1}
(-{W^0_4\over\alpha})e^y_4
(\Lambda_1 \delta(y)+\Lambda_2\delta(y-\tilde y)), \nn
\widetilde{V}_4^2&=&e^y_4(\Lambda_1 \delta(y)+\Lambda_2\delta(y-\tilde y)).
\end{eqnarray}
Elimination of these auxiliary fields by substituting these solutions 
back into the action (\ref{eq:FVinbulk}) plus (\ref{eq:braneaction}), 
could potentially yield  singular terms in the form of the squares of 
delta functions. 
However, in fact, we see that the contributions 
from $\widetilde\calF^1$ and $\widetilde\calF^2$ and from 
$\widetilde{V}_4^2$ exactly cancel each other.
%\begin{eqnarray}
%&&\hspace*{-3em}-2\bigl(1+{W^0_4^2-A^aA_a\over\alpha^2}\bigr)
%\bigl((\widetilde\calF^1)^2+(\widetilde\calF^2)^2\bigr)
%+2(\widetilde{V}_4^2)^2
%+4\left(\widetilde\calF^1-{W^0_4\over\alpha}\widetilde\calF^2
%-\widetilde{V}_4^2\right)(\Lambda_1 \delta(y)+\Lambda_2\delta(y-\tilde y)) \nn
%&=&
%\left[2\bigl(1+{W^0_4^2\over\alpha^2}\bigr)^{-1}\bigl(1+(-{W^0_4\over\alpha})^2\bigr)
%-2\right](\Lambda_1 \delta(y)+\Lambda_2\delta(y-\tilde y))^2
%\end{eqnarray}

After eliminating these auxiliary fields, the brane action becomes 
\begin{equation}
S_{\rm brane}=
\int d^5x\bigl(\Lambda_1\delta(y)+\Lambda_2\delta(y-\tilde y)\bigr)\,e_4\,\bigl(
-4g_\R Q^1\alpha 
- 2\bar\psi_{\mu}\gamma^{{\mu}{\nu}}\psi_{\nu}\bigr), 
\end{equation} 
The scalar $\alpha$ of the $U(1)_{\rm R}$ gauge multiplet is 
nonvanishing. If the $U(1)_{\rm R}$ gauging is done with $Q^1=1$, i.e.,
$t_{\rm R}=i\sigma_1$, then this essentially reproduces the brane 
action given by Altendorfer, Bagger and 
Nemeschansky.\cite{ref:ABN} The point here is, however, that the 
parameters $\Lambda_1$ and $\Lambda_2$ remain arbitrary and are not 
determined by the supersymmetry requirement at all. Therefore, despite 
the fact that the bulk cosmological constant is given by the parameter 
$g_R\alpha$, the brane tensions are $\Lambda_1$ or $\Lambda_2$ times 
$-4g_R\alpha$, and thus have no relation to the bulk cosmological 
constant $-(8/3)g^2_R\alpha^2=-4g_R^2$.\cite{ref:KO2} Zucker noted the 
same point in his off-shell Poincar\'e supergravity formulation based on
a linear multiplet compensator.\cite{ref:Zuk3}

Let us comment on the Killing spinor in the Randall-Sundrum 
background,\cite{ref:RS}
\begin{equation}
ds^2=e^{-2k|y|}\eta_{\mu\nu}dx^\mu dx^\nu-dy^2.
\end{equation}
The Killing spinor is found by demanding that the $\XQ$ and $\XS$ 
transformation $\delta=\delta_Q(\varepsilon)+\delta_S(\eta)$ of the 
gravitino $\psi^i_\mu,\ 
\psi^i_y$ and the fermion components $\Omega^{0i}$ of $\XV_0$ and 
$\zeta^a$ of $\XH^a$ vanish. The $\XS$-transformation parameter $\eta$ 
is given as a function of $\varepsilon$ by the condition $\delta 
\Omega^{0i}=0$ (and then $\delta\zeta^a=0$ is automatically satisfied). 
Assuming that the Killing spinor $\varepsilon(y)$ depends only on the extra 
dimension coordinate $y$, one can show that such a Killing spinor can 
exist only when $\Lambda_1=-
\Lambda_2=2$, $Q^1=1\ (Q^2=Q^3=0)$ and $k=2\alpha g_\R/3$. 
This implies that the brane tension 
$\pm\tau$ of the two boundary planes should be 
$\pm\tau\equiv\pm4g_\R\alpha Q^1\Lambda_1=\pm12k$ while the bulk 
cosmological term is $-4g_\R^2=-6k^2$. However, this value of the brane 
tension is twice as large as the Randall-Sundrum value, 
$\pm\tau=\pm6k^2$. Zucker also noted this fact in his 
formulation.\cite{ref:Zuk3} Since the effective four-dimensional 
cosmological term vanishes only when the Randall-Sundrum case, this 
implies that there exists no Killing spinor, and therefore the (global) 
supersymmetry is {\em spontaneously broken} on the Randall-Sundrum 
background. Note that this conclusion is in the framework of 
Altendorfer-Bagger-Nemeschansky approach. In fact, in the GPFLP approach 
whose off-shell formulation is given in Ref.~\citen{ref:FKO},
the same Randall-Sundrum background is shown\cite{ref:BKVP} to allow the
existence of a Killing spinor.

\section*{Acknowledgements}
The authors owe a lot to Tomoyuki Fujita, who collaborated with 
them at the early stages of this work. 
The authors would like to thank 
Tony Gherghetta, David E.~Kaplan, Tatsuo Kobayashi, Nobuhiro Maekawa, 
Hiroaki Nakano and Stefan Vandoren for their encouragement and interest 
in this work. They also appreciate the Summer Institute 2001 held at 
Fuji-Yoshida, the discussions at which were valuable. T.~K.\ is 
supported in part by a Grant-in-Aid for Scientific Research, No.~13640279, 
from the Japan Society for the Promotion of Science and a Grant-in-Aid 
for Scientific Research on Priority Areas, No.~12047214, from the Ministry
of Education, Science, Sports and Culture, Japan.

\appendix

\section{Conventions}

The gamma matrices $\gamma^a$ ($a=0,1,2,\cdots,d-1)$ in $d$ dimensions satisfy
$\{\gamma^a,\,\gamma^b\}=2\eta^{ab}$ and $(\gamma_a)^{\dagger}=\eta^{ab}\gamma_b$,
where $\eta^{ab}={\rm diag}(+,-,-,\cdots)$.
\,Here $\gamma^{a \cdots b}$ is the antisymmetrized product
of gamma matrices 
\begin{eqnarray}
\gamma^{a\cdots b}
=\gamma^{[a} \cdots\gamma^{b]},
\end{eqnarray}
where the square brackets $[\cdots]$ represent complete antisymmetrization 
of the indices with weight 1. Similarly, $(\cdots)$ represents complete 
symmetrization with weight 1.

In five dimensions, we choose the Dirac matrices to satisfy
\begin{eqnarray}
\gamma^{a_1\cdots a_5}=\epsilon ^{a_1\cdots a_5},
\end{eqnarray}
where $\epsilon ^{a_1\cdots a_5}$ is a
totally antisymmetric tensor with $\epsilon ^{01234}=1$.

The $SU(2)$ index $i$ ($i$=1,2) is raised and lowered with the antisymmetric 
$\epsilon _{ij}$ tensor ($\epsilon _{12}=\epsilon ^{12}=1$) according to 
the northwest-to-southeast (NW-SE) contraction convention:
\begin{eqnarray}
A^i=\epsilon ^{ij}A_j,\hspace{2em}A_i=A^j\epsilon _{ji}.
\end{eqnarray}
The charge conjugation matrix $C_5$ in 5D has the properties
\begin{eqnarray}
C_5^\T=-C_5,\hspace{2em}C_5^{\dagger}C_5=1,\hspace{2em}
C_5\gamma_aC_5^{-1}=\gamma_a^\T.
\end{eqnarray}
Our five-dimensional spinors $\psi^i$ satisfy the $SU(2)$-Majorana 
condition
\begin{eqnarray}
\bar\psi^i\equiv(\psi_i)^{\dagger}\gamma^0=\psi^{i\T}C_5
\end{eqnarray}
where spinor indices are omitted. When the $SU(2)$ indices are 
suppressed in bilinear terms of the spinors, the NW-SE contraction is 
understood, e.g. $\bar\psi\gamma^{a_1\cdots a_n}\lambda=\bar\psi^i\gamma^{a_1\cdots a_n}
\lambda_i$.

In four dimensions, the Dirac matrices satisfy
\begin{eqnarray}
\gamma^{a_1\cdots a_4}&=&-i\epsilon ^{a_1\cdots a_4}\gamma_5,\qquad 
\epsilon^{0123}=1, 
\end{eqnarray}
with the chirality matrix $\gamma_5$. 
The fifth Dirac matrix in 5D, $\gamma^4$, is anti-hermitian and 
related with $\gamma_5$ as $\gamma^4=i\gamma_5$.  
The Majorana-condition is defined by
\begin{equation}
\bar\psi\equiv\psi^\dagger\gamma^0=\psi^{\T}C_4,
\end{equation}
where the charge conjugation matrix $C_4$ in 4D has the properties
\begin{eqnarray}
C_4^{\T}=-C_4,\qquad C_4^\dagger C_4=1,\qquad C_4\gamma_aC_4^{-1}=-\gamma_a^{\T}.
\end{eqnarray}
In the main text, we take as our convention 
the relation $C_5=-C_4\gamma_5$ between 
the charge conjugation matrices in 5D and 4D.

\section{Curvatures in 4D and 5D}

The curvatures $\hatR_{\mu\nu}{}^A$ are defined by 
$\hatR_{ab}{}^A\XX_A\equiv-[\hatD_a,\,\hatD_b]$ in terms of the full 
superconformal covariant derivative 
$\hatD_a\equiv\partial_a-\sum_{A\not=P}h_a^A\XX_A$ and 
are written using the structure function in the form
\begin{equation}
\hatR_{\mu\nu}{}^A
= e_\mu^{\,b}e_\nu^{\,a}f_{ab}{}^A
=2\partial_{[\mu}^{\Span{A}}h_{\nu]}^A
+h_\nu^Ch_\mu^Bf'_{BC}{}^A\,.
\label{eq:Rexpr}
\end{equation}
Here $\XX_A$ and $h_\mu^A$ denote the transformation operators and the 
corresponding gauge fields, respectively, whose explicit contents 
in 5D and 4D are listed in Table~\ref{table:opgauge}.
The quantity $f_{AB}{}^{C}$ is a `structure function', defined
by $[\XX_{A},\XX_{B}\}=f_{AB}{}^{C}\XX_C$, 
which generally depends on the fields. The primed
$f'_{BC}{}^A$ is zero when $B=P_b$ and $C=P_c$, and otherwise 
$f'_{BC}{}^A=f_{BC}{}^A$.   
The explicit expression of (\ref{eq:Rexpr}) for the curvature 
$\hatR_{\mu\nu}{}^A$ can most easily be obtained from the gauge field 
transformation law $\delta(\varepsilon)h_\mu^{A}$ of the same generator $\XX_A$:
\begin{equation}
\varepsilon^{B}\XX_{B}\,h_\mu^{A}\equiv\delta(\varepsilon)h_\mu^{A} 
= \partial_\mu\varepsilon^{A}+\varepsilon^{C}h_\mu^{B}f_{BC}{}^{A}\,;
\end{equation}
that is, we can obtain $\hatR_{\mu\nu}{}^A$ by simply 
 replacing $\varepsilon^C$ by $h_\nu^C$ in $\delta(\varepsilon)h_\mu^{A}$.

\begin{table}[tb]
\caption{The transformation operators and the gauge fields.}
\label{table:opgauge}
\begin{center}
\begin{tabular}{l||c|c|c|c|c|c|c|c} \hline \hline
\multicolumn{9}{l}{in 5 dimensions }\\ \hline
${\XX_{A}}^{\Span{A}}$ & $\XP_a(=\hatD_a)$ & $\XQ^i$ 
& $\XM_{ab}$ & $\XD$ & $\XU_{ij}$ & $\XS^i$ & $\XK_a$ & $\XG(\XZ)$\\
${h_\mu^{A}}_{\Span{f}}$ & $e_\mu{}^a$&$\psi_\mu^i$&$\omega_\mu{}^{ab}$&$b_\mu$&$V_\mu^{ij}$&
$\phi_\mu^i$&$f_\mu{}^a$&$W_\mu$\\ \hline
\multicolumn{9}{l}{in 4 dimensions }\\ \hline
${\XX_{A}}^{\Span{A}}$ & $\XP_a(=\hatD_a)$ & $\XQ$ 
& $\XM_{ab}$ & $\XD$ & $\XA$ & $\XS$ & $\XK_a$ & $\XG$\\
${h_\mu^{A}}_{\Span{f}}$ & $e_\mu{}^a$&$\psi_\mu$&$\omega_\mu{}^{ab}$&$b_\mu$&$A_\mu$&
$\phi_\mu$&$f_\mu{}^a$&$B^g_\mu$\\ \hline
\end{tabular}
\end{center}
\end{table}
The explicit forms of the curvatures $\hatR_{\mu\nu}{}^A$ in 4D 
are given by\cite{ref:KU1,ref:vN} 
\begin{eqnarray}
\hat R_{\mu\nu}^{\ \ a}(P) &=&2\partial_{[\mu}^{\Span{a}}e^a_{\ \nu]}
-2\omega_{[\mu}^{\ ab}e_{\nu]b}+2b_{[\mu}e^a_{\ \nu]}
+2i\bar\psi_\mu\gamma^a\psi_\nu,\nn
\hat R_{\mu\nu}^{\ \ \ ab}(M) &=& 2\partial_{[\mu}^{\Span{a}}\omega_{\nu]}^{\ ab}
-2\omega_{[\mu}^{\ ac}\omega_{\nu]c}^{\ \ b}
+8f_{[\mu}{}^{[a}e_{\nu]}{}^{b]}
-4\bar\psi_{[\mu}\gamma^{ab}\phi_{\nu]} +4i\bar\psi_{[\mu}\gamma_{\nu]}\hatR^{ab}(Q),  \nn
\hat R_{\mu\nu}(D) &=& 2\partial_{[\mu}b_{\nu]}+4f_{[\mu}{}^ae_{\nu]a}
+4\bar\psi_{[\mu}\phi_{\nu]},\nn
\hat R_{\mu\nu}(A) &=& 2\partial_{[\mu}A_{\nu]}-8i\bar\psi_{[\mu}\gamma_5\phi_{\nu]},\nn
\hat R_{\mu\nu}^{\ \ \ a}(K) &=& 2\partial_{[\mu}^{\Span{a}}f^a_{\nu]}
-2\omega_{[\mu}^{\ ab}f^{\Span{a}}_{\nu]b}-2b_{[\mu}f_{\nu]}{}^a
+2i\bar\phi_\mu\gamma^a\phi_\nu+2i\bar\psi_{[\mu}\gamma_{\nu]}\hatD_b\hatR^{ab}(Q),\nn
\hat R_{\mu\nu}(Q) &=& 2\partial_{[\mu}\psi_{\nu]}
-\myfrac12\omega_{[\mu}^{\ \,ab}\gamma_{ab}\psi^{\Span{a}}_{\nu]}
+b_{[\mu}\psi_{\nu]}-\myfrac32iA_{[\mu}\gamma_5\psi_{\nu]}
+2ie_{[\mu}{}^a\gamma_a\phi_{\nu]}, \nn 
\hat R_{\mu\nu}(S) &=& 2\partial_{[\mu}\phi_{\nu]}
-\myfrac12\omega_{[\mu}^{\ \,ab}\gamma_{ab}\phi^{\Span{a}}_{\nu]}
-b_{[\mu}\phi_{\nu]}+\myfrac32iA_{[\mu}\gamma_5\phi_{\nu]}
+2if_{[\mu}{}^a\gamma_a\psi_{\nu]} \nn
&&+\myfrac{i}2\gamma^a\bigl(\widetilde{\hat R}_{[\mu a}(A)
+i\gamma_5\hat R_{[\mu a}(A)\bigr)\psi_{\nu]}\,.
%\nn
%\hat F_{\mu\nu}(B^g)&=&
%2\partial_{[\mu}^{\Span{g}}B^g_{\nu]}-g[B_\mu^g,\,B_\nu^g]
%+2i\bar\psi_{[\mu}\gamma_{\nu]}\lambda^g.
\end{eqnarray}
With the help of Bianchi identities
$[\hatD_{[a},\,[\hatD_{\Span{[}b},\,\hatD_{c]}]\,]=0$,
one can show that the constraints (\ref{eq:4DRconstraints}) 
imply the useful equalities\cite{ref:KU2}
\begin{eqnarray}
\hatR_{ab}(D)&=&-\hatR_{[ab]}(M)
=-\myfrac12\widetilde{\hatR}_{ab}(A), \nn
\widetilde{\hatR}_{ab}(Q)&\equiv&\myfrac12\epsilon _{abcd}\hatR^{cd}(Q)
=-i\gamma_5\hatR_{ab}(Q),\nn
\hatD_{[a}\hatR_{bc]}(Q)&=&-i\gamma_{[a}\hatR_{bc]}(S),\nn
\hatD^b\hatR_{ab}(Q)&=&-i\gamma^b\hatR_{ab}(S),\qquad 
\gamma\dt\hatR(S)=0,\nn
i\gamma_5\widetilde{\hatR}_{ab}(S)&\equiv&
\myfrac{i}2\epsilon _{abcd}\gamma_5\hatR^{cd}(S)
=\hatR_{ab}(S)-i\slashD\hatR_{ab}(Q),
\end{eqnarray}
where $\hatR_a{}^b(M)\equiv\hatR_{ac}{}^{cb}(M)$ and 
$\widetilde{\hatR}_{ab}\equiv(1/2)\epsilon_{abcd}\hatR^{cd}$.

The curvatures $\hatR_{\mu\nu}{}^A$  in 5D  are given explicitly 
by\cite{ref:FO,ref:BCDWHP}
\begin{eqnarray}
\hatR_{\mu\nu}{}^a(P)&=&2\partial_{[\mu}e_{\nu]}{}^a-2\omega_{[\mu}{}^{ab}e_{\nu]b}
+2b_{[\mu}e_{\nu]}{}^a+2i\bar\psi_\mu\gamma^a\psi_\nu,\nn
\hatR_{\mu\nu}{}^i(Q)&=&
2\partial^{\Span{i}}_{[\mu}\psi_{\nu]}^i-\myfrac12\omega_{[\mu}{}^{ab}\gamma_{ab}\psi_{\nu]}^i
+b_{[\mu}\psi_{\nu]}^i-2V_{[\mu}^i{}_j\psi_{\nu]}^j+\gamma_{ab[\mu}\psi_{\nu]}v^{ab}
-2\gamma_{[\mu}\phi_{\nu]}^i,\nn
\hatR_{\mu\nu}{}^{ab}(M)&=&
2\partial_{[\mu}\omega_{\nu]}-2\omega_{[\mu}{}^a{}_c\omega_{\nu]}{}^{cb}
-4i\bar\psi_{[\mu}\gamma^{ab}\phi_{\nu]}
+2i\bar\psi_{[\mu}\gamma^{abcd}\psi_{\nu]}v_{cd}\nn
&&+4i\bar\psi_{[\mu}\gamma^{[a}\hatR_{\nu]}{}^{b]}(Q)
+2i\bar\psi_{[\mu}\gamma_{\nu]}\hatR^{ab}(Q)
+8f_{[\mu}{}^{[a}e_{\nu]}{}^{b]},\nn
\hatR_{\mu\nu}(D)&=&2\partial_{[\mu}b_{\nu]}+4i\bar\psi_{[\mu}\phi_{\nu]}+4f_{[\mu\nu]},\nn
\hatR_{\mu\nu}{}^{ij}(U)&=&2\partial^{\Span{j}}_{[\mu}V_{\nu]}^{ij}
-2V_{[\mu}^i{}_kV_{\nu]}^{kj}+12i\bar\psi_{[\mu}^{(i}\phi_{\nu]}^{j)}
-4i\bar\psi_{[\mu}^i\gamma\dt v\psi_{\nu]}^j
+\myfrac{i}2\bar\psi_{[\mu}^{(i}\gamma^{\Span{(i}}_{\nu]}\chi^{j)}_{\Span{[\mu}}\,,\nn
\hatR_{\mu\nu}{}^i(S)&=&2\partial^{\Span{i}}_{[\mu}\phi_{\nu]}^i
-\myfrac12\omega_{[\mu}{}^{ab}\gamma_{ab}\phi_{\nu]}^i
-b_{[\mu}\phi_{\nu]}^i-2V_{[\mu}^i{}_j\phi_{\nu]}^j+2f_{[\mu}^{\ \,a}\gamma_a\psi_{\nu]}^i
+\cdots,\nn
\hatR_{\mu\nu}{}^a(K)&=&2\partial_{[\mu}f_{\nu]}{}^a-2\omega_{[\mu}{}^{ab}f_{\nu]b}
-2b_{[\mu}f_{\nu]}{}^a+2i\bar\phi_\mu\gamma^a\phi_\nu+\cdots\,.
%\nn
%\hat F_{\mu\nu}(W)&=&2\partial_{[\mu}W_{\nu]}-g[W_\mu,W_\nu]
%+4i\bar\psi_{[\mu}\gamma_{\nu]}\Omega-2i\bar\psi_\mu\psi_\nu M.
\end{eqnarray}
Here, the dots in the $\XS^i$ and $\XK^a$ curvature expressions 
denote terms containing the other curvatures. 
The constraints (\ref{eq:5DRconstraints}) in 5D also imply 
the equalities, 
\begin{eqnarray}
\hatR_{ab}(D)&=&-\myfrac23\hatR_{ab}(M)=0,\qquad 
\hatR_{[abc]d}(M)=\hatR_{a[bcd]}(M)=0,\nn
\hatR_{ab}{}^i(S)&=&\slashD \hatR_{ab}{}^i(Q)
+\gamma_{[a}\hatD^c\hatR_{b]c}{}^i(Q)+\hatR_{[a}{}^{c\,i}(Q)v_{b]c} \nn
&&\quad {}
+\myfrac14\gamma\dt v\hatR_{ab}{}^i(Q)
+\myfrac1{12}\gamma_{ab}\hatR_{cd}{}^i(Q) v^{cd}\,,\nn
\hatR_{ab}{}^c(K)&=&
\myfrac14\hatD_d\hatR_{ab}{}^{cd}(M)
+\myfrac12\bar{\hatR}_{d[a}(Q)\gamma^c\hatR^d{}_{b]}(Q)
+\myfrac12\bar{\hatR}_{d[a}(Q)\gamma_{b]}\hatR^{dc}(Q)\,,
\hspace{2em}
\end{eqnarray}
and the $\XS^i$ and $\XK^a$ curvatures 
can be written in terms of the other curvatures.
\section{Embedding and Invariant Action Formulas in 4D}

A product of chiral multiplets also forms a chiral multiplet.
More generally, for an arbitrary set of chiral multiplets $\Sigma^I
=[\calA^I,\ \calP\chi^I,\ \calF^I]$, we can have a new chiral multiplet 
${g}(\Sigma)$, whose first component is given by a general function 
${g}(\calA)$ of $\{\calA^I\}$ carrying a homogeneous degree in the 
Weyl weight:\cite{ref:KU1}
\begin{eqnarray}
{g}(\Sigma)=[{g}(\calA),\ \calP_\R\chi^I{g}_I(\calA),\ 
\calF^I{g}_I(\calA)
-\myfrac14\bar\chi^I\calP_\R\chi^J{g}_{IJ}(\calA)],
\end{eqnarray}
where ${g}_I(\calA)\equiv\partial{g}(\calA)/\partial\calA^I$ and 
${g}_{IJ}(\calA)\equiv\partial^2{g}(\calA)/\partial\calA^I\partial\calA^J$. 
%Although we mentioned in the text only to the {\em scalar} 
%chiral multiplet that carries no Lorentz spinor indices, 
%the gauge multiplet $[B_\mu^g,\lambda^g,D^g]$ can in fact be embedded into a 
%{\em spinor} chiral multiplet $\Sigma^\alpha$ of $w=n=3/2$ carrying a 
%spinor index $\alpha$;
%\begin{eqnarray}
%\Sigma^\alpha&=&\left[(\bar\lambda^g\calP_\R)^\alpha,\,
%\left((\myfrac12\gamma\dt \hat F(B^g)+iD^g{\bf 1})\calP_+\right)_\beta{}^\alpha, \,
%(i\hatD_a\bar\lambda^g\gamma^a\calP_\R)^\alpha\right].
%\end{eqnarray}

Similarly, for an arbitrary set of general multiplets $\Phi^I=[C^I,\ \zeta 
^I,\ \cdots]$ and an arbitrary function $f(C)$ homogeneous 
in the Weyl weight,
we can have a new general multiplet $\Phi'=f(\Phi)$ whose components are 
given by\cite{ref:KU2}
\begin{eqnarray}
C'&=&f(C), \hspace{7.8em} \zeta'=\zeta^If_I,\nn
H'&=& H^If_I-\myfrac14\bar\zeta^{I}\zeta^Jf_{IJ},\qquad 
K'=K^If_I+\myfrac14\bar\zeta^Ii\gamma_5\zeta Jf_{IJ},\nn
B'_a&=&B_a^If_I+\myfrac14\bar\zeta^I\gamma_a\gamma_5\zeta^Jf_{IJ},\nn
\lambda'&=&\lambda^If_I-\myfrac12i\gamma_5\left(i\gamma_5H+K+i\slash B+\slashD C\gamma_5
\right)^I\zeta^Jf_{IJ}-\myfrac14\zeta^I\bar\zeta^J\zeta^Kf_{IJK},\nn
D'&=&D^If_I+\myfrac12(H^IH^J+K^IK^J+B^I_aB^{aJ}+\hatD_aC^I\hatD^aC^J
+i\bar\zeta^I\slashD\zeta^J-2\bar\zeta^I\lambda^J)f_{IJ}\nn
&& +\myfrac14\bar\zeta^Ii\gamma_5(i\gamma_5H+K+i\slash B)^J\zeta^Kf_{IJK}+\myfrac1{16}\bar\zeta^I\zeta^J\bar\zeta^K\zeta^Lf_{IJKL},
\end{eqnarray}
with $f_I\equiv\partial f(C)/\partial C^I$, etc. This formula is also valid for
complex general multiplets, provided that the spinor 
conjugate $\bar\psi$ is understood to be $\psi^\T C_4$, not $\psi^\dagger\gamma^0$.

For a chiral multiplet 
$\Sigma^{(w=3)}=[\calA,\ \calP_\R\chi,\ \calF]$ 
with weight $w=n=3$, we have the following 
superconformal-invariant $F$-term action formula:\cite{ref:KTvN}
\begin{eqnarray}
I_F=\int d^4x\bigl[\Sigma^{(w=3)}\bigr]_F
=\int d^4x\,e\left[\calF-i\bar\psi\dt\gamma\calP_\R\chi 
-2\bar\psi_a\gamma^{ab}\calP_\L\psi_b\calA+\hbox{h.c.}\right].\qquad 
\hspace{1em}%&=&F-i\bar\psi\dt\gamma\chi-\bar\psi_a\gamma^{ab}(A-i\gamma_5B)\psi_b.
\label{eq:Fformula}
\end{eqnarray}
For a real general multiplet 
$\Phi^{(w=2,\,n=0)}=[C,\ \zeta,\ H,\ K,\ B_a,\ \lambda,\ D]$ 
with weight $w=2,\,n=0$, 
the invariant $D$-term action formula is given by\cite{ref:KU1}
\begin{eqnarray}
I_D&=&\int d^4x\bigl[\Phi^{(w=2,\,n=0)}\bigr]_D\nn
&=&\int d^4x\,e\left[D-\bar\psi\dt \gamma\gamma _5\lambda 
+i\epsilon^{abcd}\bar\psi_a\gamma_b\psi_c\left(B_d-\bar\psi_d\zeta\right)\right.\nn
&&\left.\qquad \quad 
+\myfrac13(R(\omega)+4i\bar\psi_\mu\gamma^{\mu\nu\lambda}\calD_\nu\psi_\lambda)C
+\myfrac23i\bar\zeta\gamma_5\gamma^{\mu\nu}\calD_\mu\psi_\nu\right]\,,
\label{eq:Dformula}
\end{eqnarray}
where $R(\omega)$ is the scalar curvature constructed from the spin connection 
$\omega_\mu^{ab}$, and $\calD_\mu$ is the covariant derivative with respect to 
the homogeneous transformations $\XM_{ab},\ \XD$ and $\XA$.

\end{document}